\documentclass[12pt,a4paper]{article}

\usepackage{epsfig,multicol,multirow,cite}
\usepackage{amsmath,subfigure,latexsym,amssymb}
\usepackage[dvips]{color}

\def\beq{\begin{equation}}              
\def\eeq{\end{equation}}      
\def\bea{\begin{eqnarray}}               
\def\eea{\end{eqnarray}}           

\def\lsi{\raise0.3ex\hbox{$<$\kern-0.75em\raise-1.1ex\hbox{$\sim$}}}
\def\gsi{\raise0.3ex\hbox{$>$\kern-0.75em\raise-1.1ex\hbox{$\sim$}}}

\def\backder{\raise1.4ex\hbox{$\leftarrow$\kern-0.75em\raise-1.4ex\hbox{$\partial$}}}

\def\forder{\raise1.4ex\hbox{$\rightarrow$\kern-0.75em\raise-1.4ex\hbox{\!\!$\partial$}}}

\newcommand{\lsim}{\mathop{\lsi}}

\newcommand{\backderi}{\mathop{\backder}}
\newcommand{\forderi}{\mathop{\forder}}

\newcommand{\nn}{\nonumber}
\newcommand{\NN}{{\kern+.25em\sf{N}\kern-.78em\sf{I} \kern+.78em\kern-.25em}}

\begin{document}
{
\begin{flushright}
{\small HU-EP-07/09} \\  
{\small DESY-07-043} \\
{\small MIT-CTP 3827}
\end{flushright}
}

\vspace*{3mm}

\begin{center}     
{\large\bf Area-preserving diffeomorphisms in gauge theory}
\vspace*{3mm} \\
{\large\bf on a non-commutative plane: a lattice study}
\end{center}
\vspace{1ex}
\begin{center}
Wolfgang Bietenholz$^{\rm \, a}$, Antonio Bigarini$^{\rm \, b,c}$
and Alessandro Torrielli$^{\rm \, d}$ 
\end{center}


\small

\begin{center}
$^{\rm a}$ John von Neumann Institut f\"{u}r Computing (NIC) \\ 
Deutsches Elektron Sychrotron (DESY) \\
Platanenallee 6, D-15738 Zeuthen, Germany \\

\vspace{4mm}

$^{\rm b}$ Dipartimento di Fisica, Universit\`{a} degli Studi di Perugia\\
and INFN, Sezione di Perugia, \\
Via Pascoli 1, I-06100 Perugia, Italy \\ 

\vspace{4mm}

$^{\rm c}$ Institut f\"{u}r Physik \\
Humboldt-Universit\"{a}t zu Berlin \\
Newtonstr.\ 15, D-12489 Berlin, Germany \\

\vspace{4mm}

$^{\rm d}$ Center for Theoretical Physics, Laboratory for Nuclear Sciences \\
and Department of Physics \\ 
Massachusetts Institute of Technology \\
77 Massachusetts Avenue, Cambridge, MA 02139-4307, USA
\end{center}
\vspace{4ex}

\normalsize 

\noindent
We consider Yang-Mills theory with the $U(1)$ gauge group
on a non-commu\-ta\-tive plane.
Perturbatively it was observed that the invariance of this theory under
area-preserving diffeomorphisms (APDs) breaks down to a 
rigid subgroup $SL(2,R)$. 
Here we present explicit results for the APD symmetry breaking at finite
gauge coupling and finite non-commuta\-ti\-vity. They are based
on lattice simulations and measurements of Wilson loops with 
the same area but with a variety of different shapes. 
Our results are consistent with the expected loss of invariance under APDs.
Moreover, they strongly suggest that non-perturbatively the
$SL(2,R)$ symmetry does not persist either. 

\newpage

\section{Introduction}

Invariance under area-preserving diffeomorphisms (APDs) 
\cite{witten} is a basic symmetry of ordinary Yang-Mills
theories in two dimensions. 
In particular it means that Wilson loop expectation
values only depend on the oriented areas singled out on the manifold.
Thanks to this property the 
theory acquires an almost topological flavour \cite{mig} and, as a 
consequence, it can be solved analytically. Elegant 
group theoretic methods \cite{group} lead to 
closed expressions for the partition function and a set of
observables \cite{migmak}. 

The invariance under APDs was initially believed to persist
also in $U(n)$ gauge theories defined on a non-commutative (NC)
two-dimensional mani\-fold. It was assumed to play a 
central role in the large gauge group --- characteristic of 
gauge theories on NC spaces --- which merges 
internal and space-time transformations. 
A detailed study of the non-commutative 
gauge-transformation algebra was performed in Ref.\ \cite{LSZ}. 
If APD symmetry holds, one might hope
to be able to solve gauge theories also on a NC plane by
generalising the powerful geometric procedures developed in the 
commutative space.

This scenario was suggested by an intriguing observation
for $U(n)$ gauge theory on a NC torus. It can be related by 
Morita equi\-valence to its dual on a
commutative torus \cite{MMSJ}, 
where the APD invariance is granted. 
The theory on the NC plane would then be reached by a suitable limit,
and one could hope for the invariance to be preserved \cite{ps2}.   

Wilson loop perturbative expansions in the coupling constant $g$ and 
in $1/{\theta}$ --- $\theta$ being the non-commutativity parameter --- 
were performed on the NC plane. To the order first considered 
in Refs.\ \cite{bnt1,bnt2}, the results were consistent 
with APD invariance. 

Later on Ref.\ \cite{adm} extended
those results to the next order, namely ${\cal O}(\theta^{-2})$ 
at ${\cal O}(g^4)$. The outcome revealed {\em different} expectation
values for a Wilson loop with the shape of a circle and a rectangle
of the same area. This observation motivated the systematic investigation 
in Ref.\ \cite{noi}, where Wilson loops in a wide class of contours 
were considered in the axial gauge. These results
suggest that the APD symmetry breaks down to a residual subgroup
of linear unimodular transformations, $SL(2,R)$.

Subsequently non-perturbative arguments
for this APD symmetry breaking were given
based on the Morita duality on tori \cite{CGSS}.
Recently, Ref.\ \cite{RicSza} reconsidered this issue by
applying twist deformation techniques, which also confirm that
the APD symmetry may break at the quantum level.
A new study of the large $\theta$ expansion \cite{ADM2}
reports the break-down of the area law at ${\cal O}(1 / \theta^{2})$, 
in agreement with simulations results at finite $\theta$ \cite{NCQED2}.
Ref.\ \cite{ADM2} does not report the observation of any symmetry.
However, the question about the ultimate status of (partial)
APD symmetry is still open.

The present work presents explicit results for
Wilson loops with polygonal contours, at finite $\theta$ and $g$, 
under APDs. Our non-perturbative results are obtained on the lattice
and extrapolated to the continuum. They agree with the breaking 
of this symmetry, both on the lattice and in the continuum limit.
Moreover we provide evidence against 
the survival of a residual symmetry subgroup $SL(2,R)$.

In Section 2 we briefly review the $U(1)$ gauge theory on a NC plane,
its lattice discretisation and the mapping onto a twisted
Eguchi-Kawai model, which can be simulated.
Section 3 presents our simulation results for
the planar limit, which is necessary to identify a physical scale. 
Then we address in Section 4 the Double Scaling Limit
to a continuous NC plane of infinite extent, which allows us to
study explicitly the effect of APD transformations on Wilson 
loops. In Section 5 we confirm the APD symmetry breaking by 
considering observables, which differ from those in Section
4. Their perturbative treatment is commented on in an Appendix.
Section 6 focuses specifically on the $SL(2,R)$ symmetry, and
Section 7 is dedicated to our conclusions.

\section{$U(1)$ gauge theory on a non-commutative plane}

In this work, we consider the simplest version of a Euclidean NC plane
by assuming a constant non-commutativity parameter $\theta$, so that
the coordinate operators fulfil
\beq
[ \hat x_{\mu}, \hat x_{\nu} ] = {\rm i} \, \theta \, 
\epsilon_{\mu \nu} \qquad
(\mu , \nu = 1,2 ) \ .
\eeq
Such coordinates describe a charged particle moving in a 
(commutative) plane, which is crossed by a strong, orthogonal 
magnetic field. The latter can be formally interpreted as 
$B \propto 1 / \theta$, see e.g.\ Ref.\ \cite{NCmag}.
A similar concept is also used to map open strings 
in a magnetic background onto NC field theory \cite{SeiWit}.

We can return to the use of ordinary (commutative) coordinates if all the 
fields are multiplied by {\em star products} (or {\em Moyal products}),
\beq
\phi (x) \star \psi (x) := \phi (x) \exp \Big( \, 
\frac{\rm i}{2} \backderi \,\! _{\! \mu} \, \theta \, \epsilon_{\mu \nu}
\forderi \,\! _{\!\! \nu} \, \Big) \ \psi (x) \ .
\eeq
Here we focus on pure $U(1)$ gauge theory with the Euclidean action
\bea
S[A] &=& \frac{1}{4} \int d^{\, 2}x \, F_{\mu \nu} \star F_{\mu \nu} \ , 
\nn \\
F_{\mu\nu} &=& \partial_{\mu} A_{\nu} - \partial_{\nu} A_{\mu}
+ ig [ A_{\mu},A_{\nu}]_{\star} \ .
\eea
The last term is a star-commutator, which shows that
even the $U(1)$ gauge field is self-interacting on NC spaces.
This action is star-gauge invariant, i.e.\ invariant under transformations
\beq
A_{\mu}(x) \to U(x) \star A_{\mu}(x) \star U(x)^{\dagger}
- \frac{\rm i}{g} U(x) \star \partial_{\mu}U(x)^{\dagger} \ ,
\eeq
if $U(x)$ is star-unitary, $U(x)^{\dagger} \star U(x) = 1 \!\! 1$.

Other $U(n)$ gauge theories 
may be studied along the same lines, but the formulation of $SU(n)$
gauge theories runs into trouble on NC spaces. 
Therefore it is motivated to concentrate on $U(1)$ as a physical gauge group,
which can be accommodated on NC manifolds. \\

Although the points in such spaces are somewhat fuzzy,
it is possible to introduce a lattice structure.\footnote{Here 
we only sketch this regularisation very briefly,
for details we refer for instance to the review \cite{SzaboRev},
or the theses quoted in Refs.\ \cite{NCQED2} and \cite{NCQED4}.}
This is a first step towards a formulation to be used in Monte Carlo
simulations. In the operator formalism this step imposes the constraint
\beq
\exp \Big( {\rm i} \frac{2 \pi}{a} \hat x_{\mu} \Big) = \hat 1 \!\! 1 \ ,
\eeq
where $a$ is the lattice spacing.
If we require the momentum components to be commutative and periodic 
over the Brillouin zone, the above condition implies that only discrete 
momenta occur, which is characteristic for a finite volume.
On a $N \times N$ lattice with periodic boundary conditions, 
the allowed momenta are spaced by \ $ 2\pi / (aN) $. As a consequence, 
the non-commutativity parameter can be identified as
\beq
\theta = \frac{1}{\pi}N a^{2} \ .
\eeq
We are most interested in a {\em Double Scaling Limit} (DSL)
\beq  \label{DSLeq}
a \to 0 \quad {\rm and} \quad N \to \infty \qquad
{\rm at} \qquad  Na^{2} = const. \ ,
\eeq
which leads to a continuous NC plane of infinite extent. The requirement
to take the UV and IR limits simultaneously in a balanced way is
related to the generic UV/IR mixing of the divergences in NC
field theory \cite{UVIR}.

This is clearly distinct from the {\em planar limit,} $N \to \infty$ at 
fixed gauge coupling, which means here a fixed lattice spacing. 
The non-commutativity parameter
diverges in this limit. In higher dimensions, this implies that non-planar
contributions are suppressed, and the planar limit restores
commutativity in perturbation theory.\footnote{We remark, however, that
this restoration does not need to hold generally: it can fail
non-perturbatively for instance in the case of spontaneous symmetry breaking 
\cite{NCplanar}.} In two dimensions the situation is different, and 
non-planar diagrams provide ``anomalous'' perturbative 
contributions in the limit of infinite non-commutativity, 
which are of the same order of magnitude as the planar diagrams
\cite{bnt1,bnt2,adm}.
However, such terms are shown to disappear when applying the 
procedure recently introduced in Ref.\ \cite{ADM2}. \\
 
Even on the lattice it is far from obvious how to simulate NC 
gauge theory; note that the compact formulation seems to require
star-unitary link variables. In this respect, it is highly 
profitable to map the system onto a {\em twisted Eguchi-Kawai model}
(TEK model). This model is defined on a single 
space-point and its action takes the form \cite{TEK}
\beq  \label{TEKact}
S_{\rm TEK} [U] = - N \beta \sum_{\mu \neq \nu} Z_{\mu \nu} {\rm Tr} \,
\Big( U_{\mu} U_{\nu} U_{\mu}^{\dagger} U_{\nu}^{\dagger} \Big) \ .
\eeq
$U_{1}$ and  $U_{2}$ are unitary $N \times N$ matrices which encode 
the degrees of freedom of the $U(1)$ lattice gauge theory. 
For the twist factor we adopt the choice of Ref.\ \cite{NCQED2},
$\, Z_{21} = Z_{12}^{*} = \exp ( {\rm i} \pi (N+1)/N) \, $, where $N$ has to 
be odd. There is an exact equivalence to the lattice NC $U(1)$ gauge theory,
i.e.\ the algebras are identical, as Ref.\ \cite{AIIKKT} showed
in the large $N$ limit. A refined consideration found such a mapping
even at finite $N$ \cite{AMNS}. Hence the TEK model can be used
for numerical simulations of NC gauge theories, and it is
most suitable for this purpose.

It is straightforward to formulate 
Wilson loops in this matrix model.
For instance, a rectangular loop with side lengths $aI$ and $aJ$
(and clockwise orientation) corresponds to the term
\beq
W ( I \times J) = \frac{1}{N} Z_{12}^{IJ} \
{\rm Tr} \, \Big( U_{1}^{I} U_{2}^{J} U_{1}^{\dagger \, I} 
U_{2}^{\dagger \, J} \Big) \ .
\eeq
Mapping this quantity back to the lattice leads in fact to
a sensible definition of a Wilson loop in the NC gauge theory 
\cite{IIKK,Wloop}.
Such Wilson loops are complex in general \cite{NCQED2,bnt2}. 
The action (\ref{TEKact}) is real, however, since both
orientations of the plaquettes are summed over, 
which is essential for the feasibility of numerical simulations.

Simulations of gauge theories with the standard Metropolis
algorithm are notoriously inefficient. This also holds for the
TEK model. Moreover the usual remedy --- the application of the 
heat-bath algorithm --- cannot be applied straightforwardly,
because the dimensionally reduced action (\ref{TEKact}) is non-linear
in the link variables. However, by introducing an auxiliary
matrix field the action can be linearised \cite{Fab84}, so that the
heat-bath algorithm works and the model can indeed be simulated
efficiently.\footnote{This method has recently been
extended to the 4d model where two dimensions are reduced; 
this is adequate for QED in a four dimensional space, composed of
a commutative and an NC plane \cite{NCQED4}.}
This method allowed us to explore rather large systems of $N > 100$.
For the parameter sets $(N, \beta)$ that we investigated,
we collected statistics of $1000$ well thermalised and decorrelated
configurations.

\section{The planar limit}

In the planar limit we obtain the $U( N \to \infty )$ 
lattice gauge theory on a commutative plane,
which was solved by Gross and Witten 
\cite{GroWit}. In this limit they found an exact area law for the 
Wilson loops. In (dimensionless) lattice units it takes the form
\beq  \label{arealaw}
\langle W ( I \times J) \rangle =
\exp (- \sigma (\beta ) IJ ) \ , \quad
\sigma (\beta ) = \left\{ \begin{array}{ccc}
- \ln \beta && \beta \leq 1/2 \\
- \ln (1 - \frac{1}{4 \beta}) && \beta \geq 1/2
\end{array} \right. \ .
\eeq
In terms of dimensional units the string tension $\sigma$ 
turns into an area, and it allows us therefore to identify a
dimensional lattice spacing as
\beq  \label{adim}
a = \sqrt{\sigma (\beta )} \ .
\eeq
In these units the string tension is
set to $1$ in the planar limit.

Of course, we are ultimately interested in the DSL according
to eq.\ (\ref{DSLeq}). But to give it an explicit meaning
we first have to identify a dimensional lattice spacing,
i.e.\ we have to introduce a scale to interpret the lattice units.
Relation (\ref{adim}) can be used for this purpose \cite{NN,NCQED2},
provided that the values of $N$, which are accessible
to our simulations, do approximate the planar limit well
(for the quantities of interest).
Otherwise one would have to worry about finite $N$ artifacts
distorting the physical interpretation of our results.

\begin{figure}
  \includegraphics[width=1\linewidth]{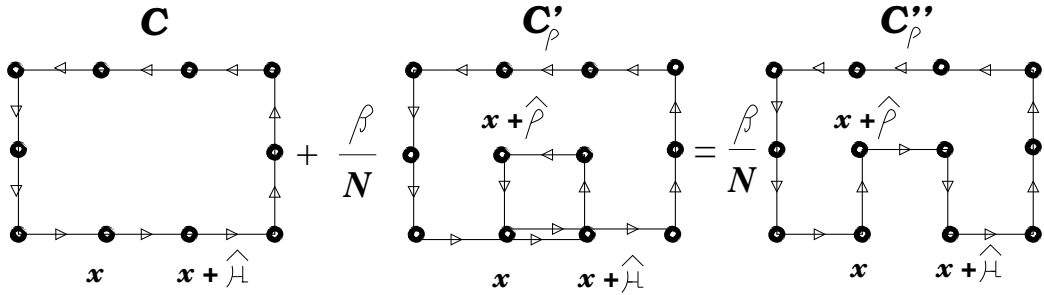}
\caption{\emph{An example of a set of contours, 
which are involved in a Schwinger-Dyson equation. 
These equations relate the vacuum expectation values 
of the corresponding Wilson loops in the planar limit.
They can be derived from the invariance under an infinitesimal
substitution of the compact link variables on the lattice \cite{TEK}.}}
\label{SDcont}
\end{figure}

\begin{figure}[ht!]
\hspace*{-2mm}
 \includegraphics[width=.36\linewidth,angle=270]{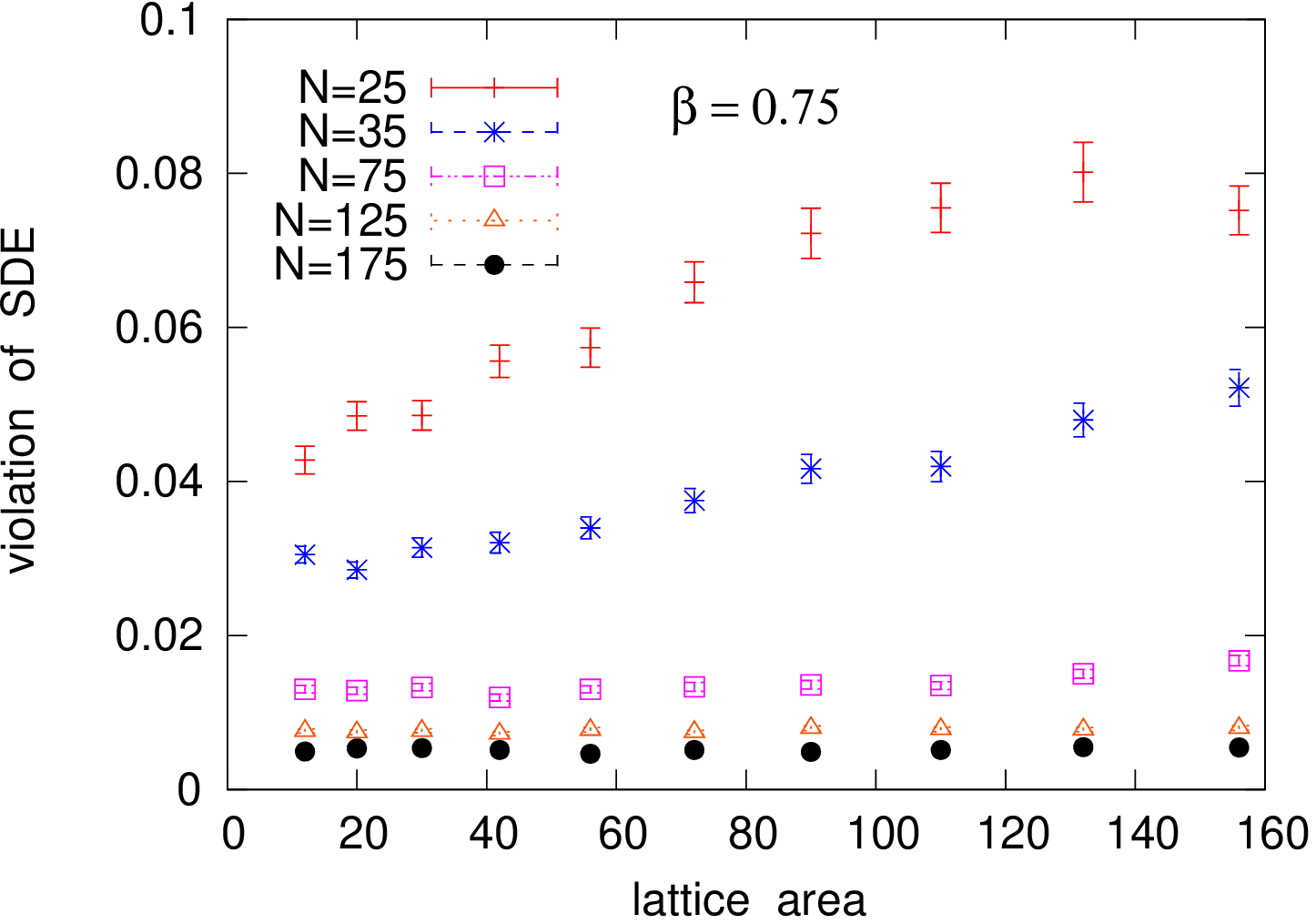}  
\hspace*{-2mm}
 \includegraphics[width=.36\linewidth,angle=270]{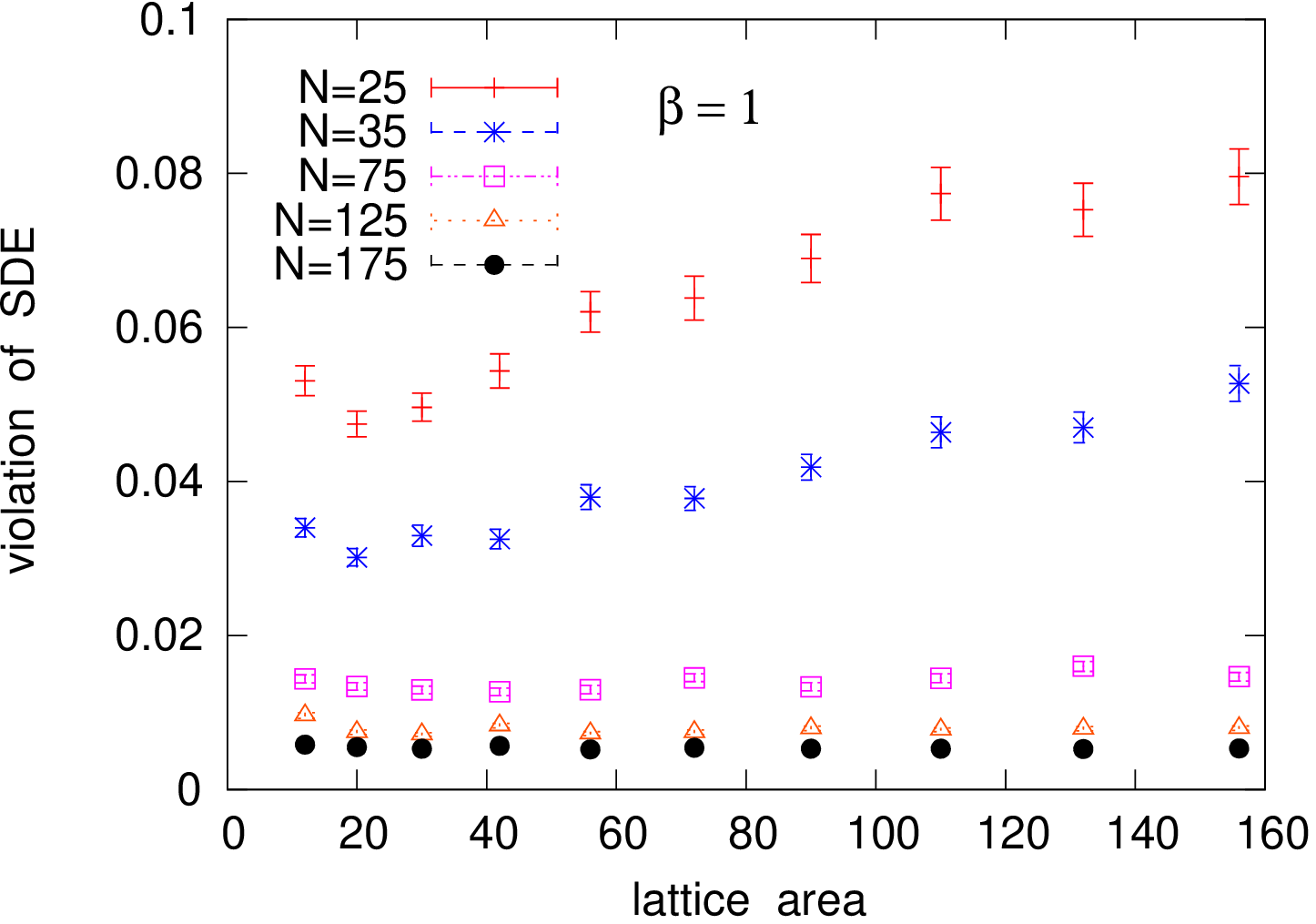}
\vspace*{-3mm}
  \caption{\emph{The convergence towards the validity of the Schwinger-Dyson 
equations as $N$ increases at fixed $\beta = 0.75$ (on the left)
and $\beta = 1$ (on the right). 
The $y$-axis is the absolute value of the
deviation between the two sides of the specific Schwinger-Dyson 
equation illustrated in Figure \ref{SDcont}.
The $x$-axis corresponds to the lattice area of the contour $C$,
which has a rectangular shape of the form $I \times (I+1)$.
These equations are well approximated as $N$ reaches ${\cal O}(100)$.}}
\label{SDfig}
\end{figure}

We verified this property first by checking
the validity of the large $N$ Schwinger-Dyson equations. 
On the lattice, they relate Wilson loops of different shapes in 
the planar limit.\footnote{This was the property that motivated the 
original construction of the dimensionally reduced matrix model
(without twist factor) by Eguchi and Kawai \cite{EK}.} 
An example for the corresponding contours is illustrated 
in Figure \ref{SDcont} (cf.\ first work in Ref.\ \cite{TEK}).
Indeed, we observe that our measurements for the two sides of this
equation converge well as we increase $N$ to a magnitude of
${\cal O}(100)$ at fixed $\beta $, see Figure \ref{SDfig}.\footnote{A 
variant of this result has been reported before in a proceeding contribution
\cite{Ahren04} and in a Ph.D.\ thesis \cite{Antonio}. Earlier observations
in this context were given in Ref.\ \cite{NN}.} 

\begin{figure}[ht!]
\begin{center}
 \includegraphics[width=.25\linewidth,angle=0]{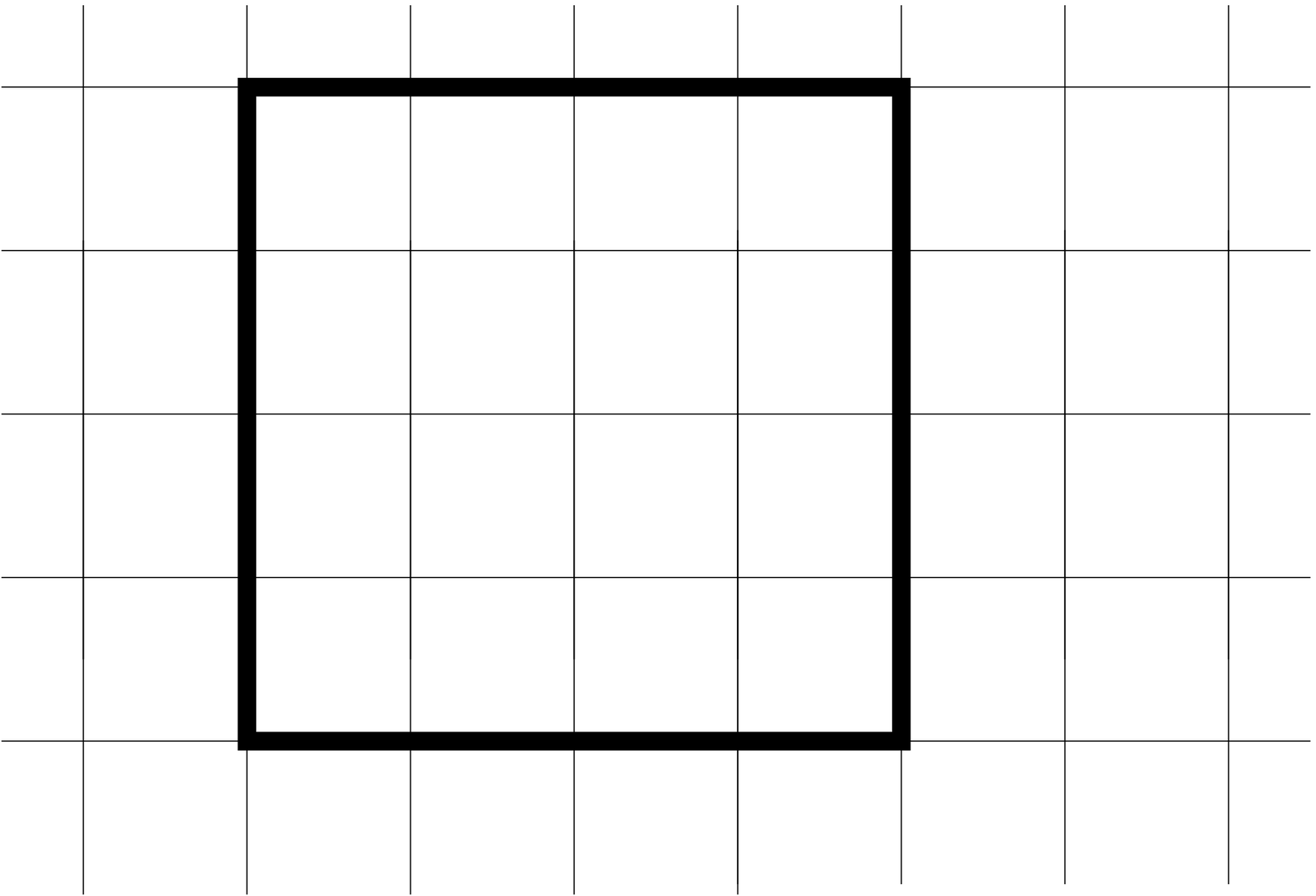} 
\hspace*{3mm}
 \includegraphics[width=.25\linewidth,angle=0]{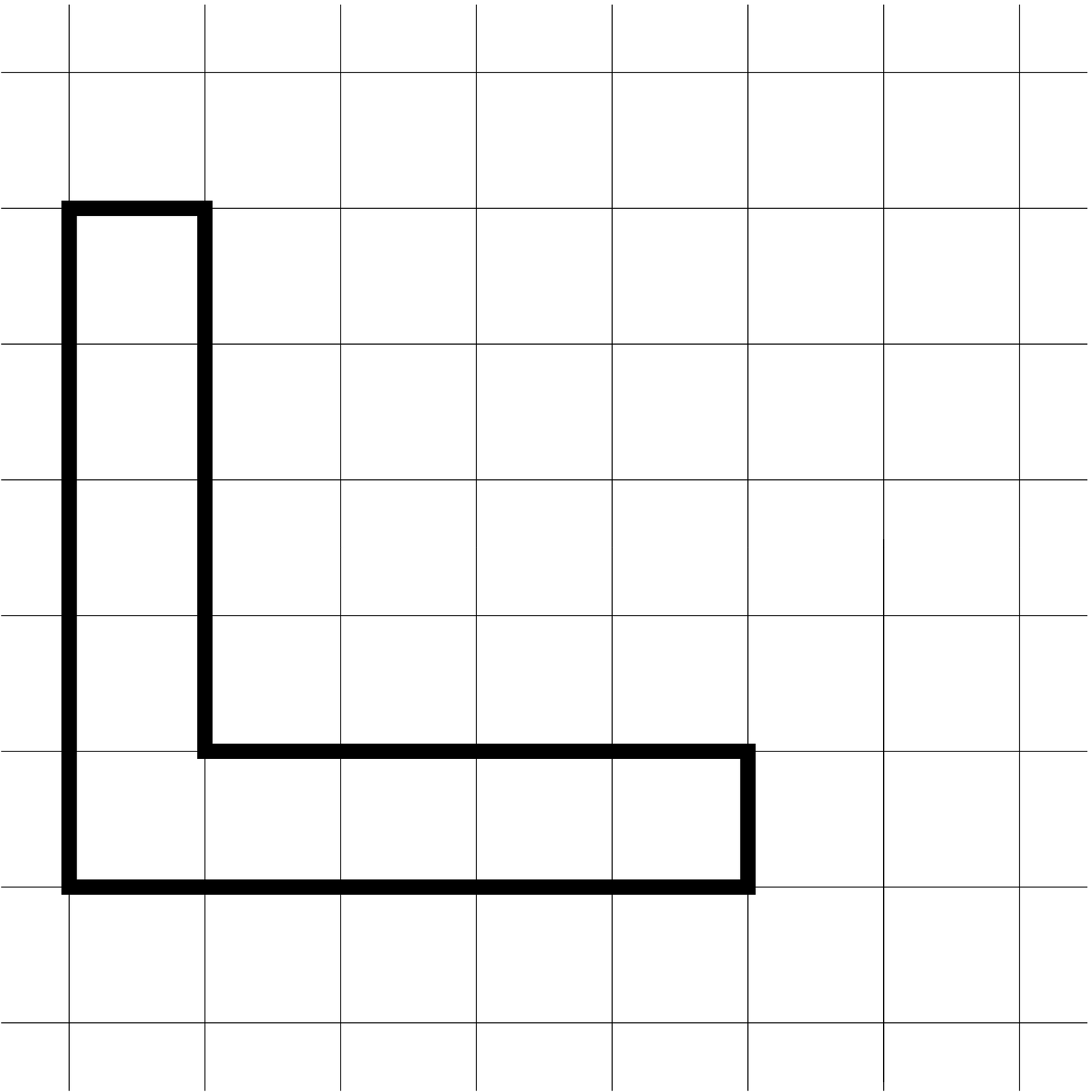} 
\hspace*{3mm}
 \includegraphics[width=.25\linewidth,angle=0]{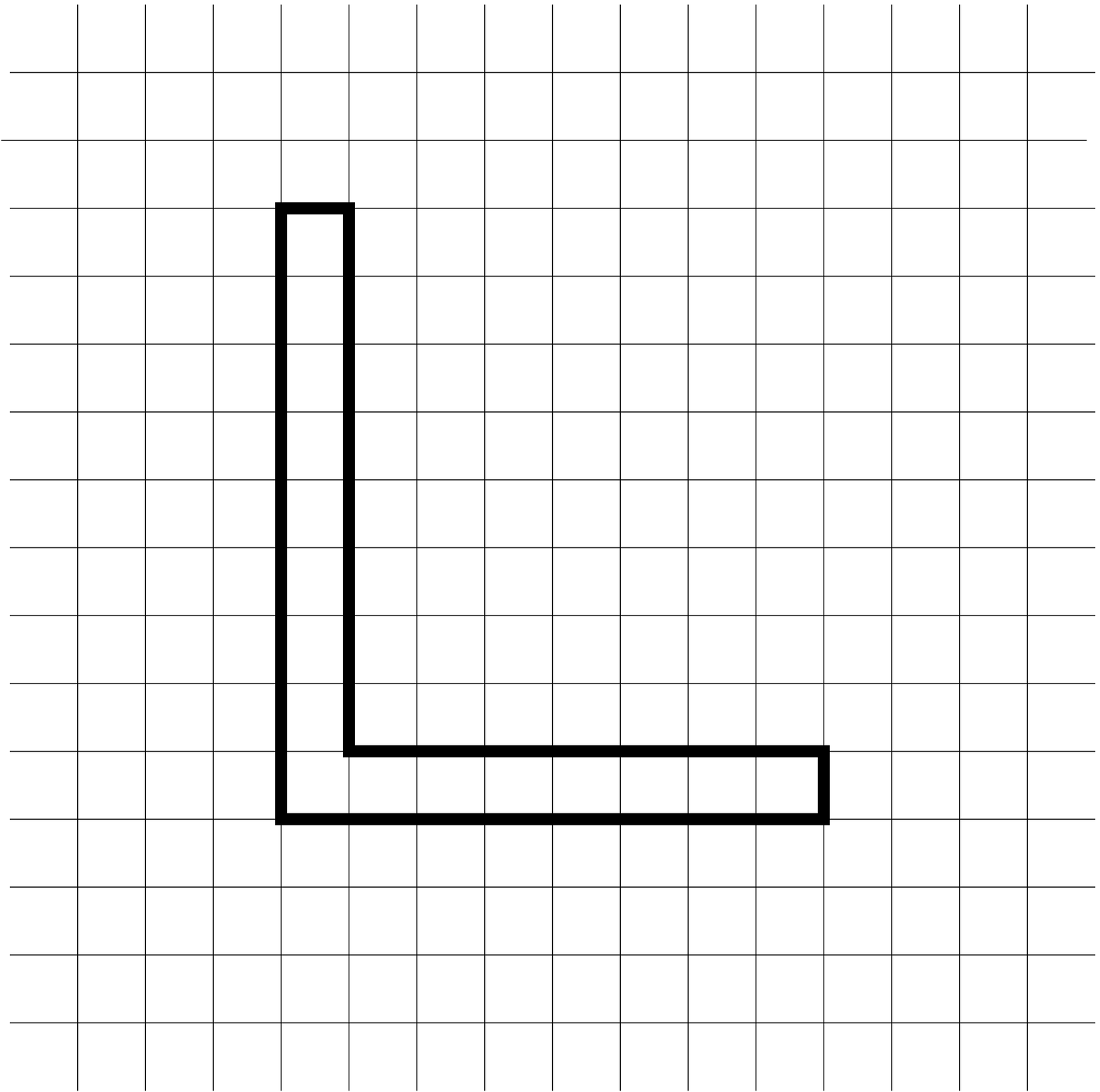} \\
\vspace*{3mm}
  \includegraphics[width=.25\linewidth,angle=0]{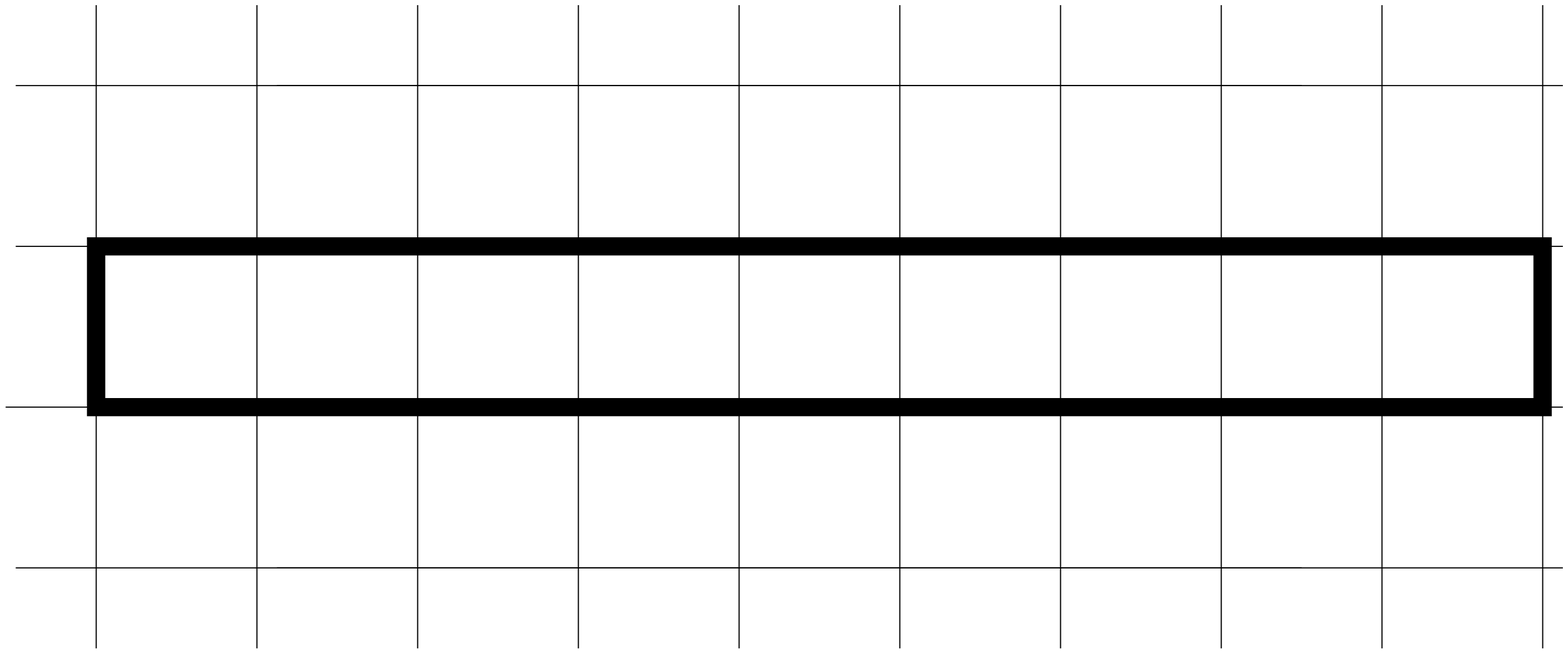} 
\hspace*{3mm}
 \includegraphics[width=.25\linewidth,angle=0]{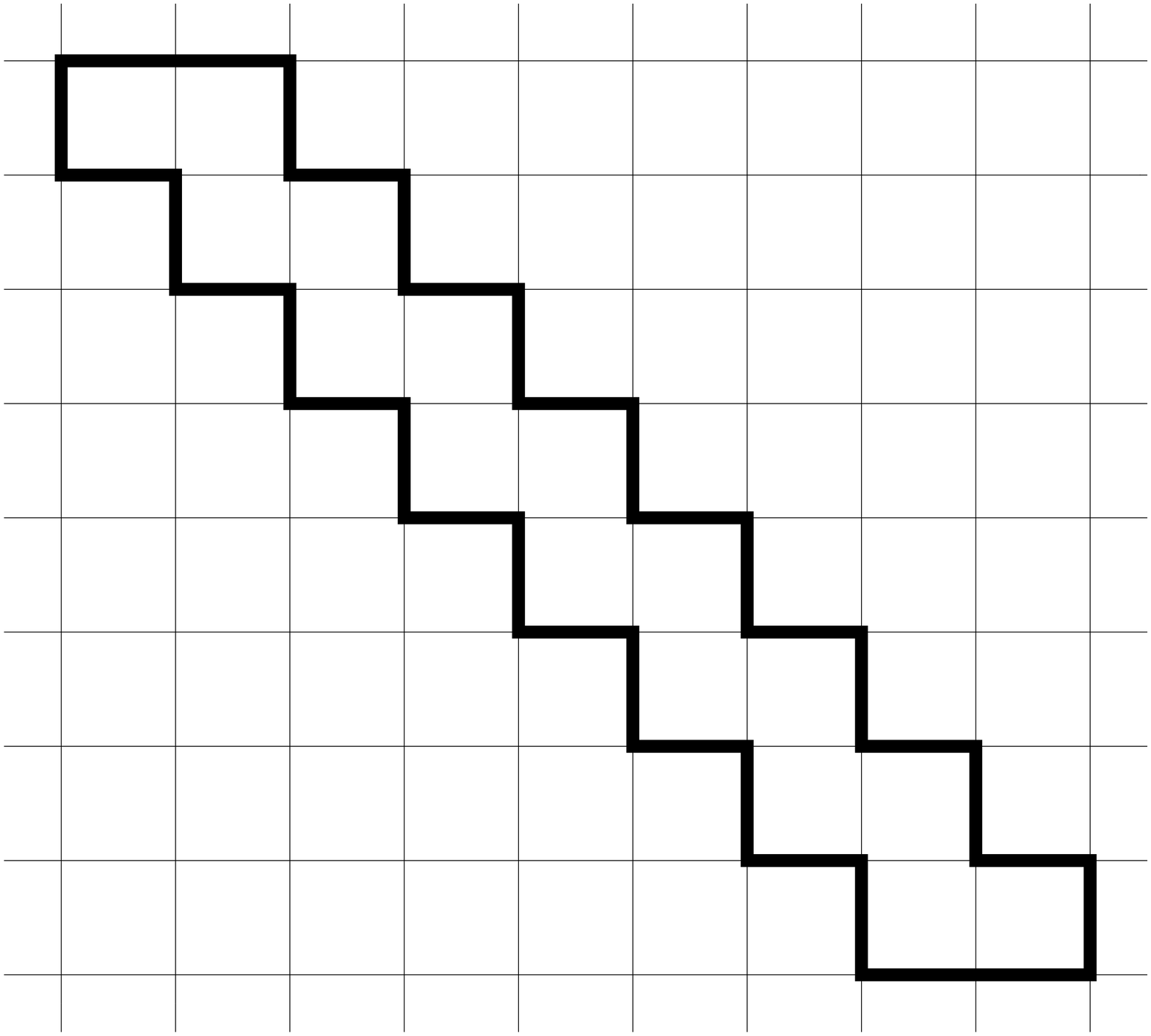}
\hspace*{3mm}
 \includegraphics[width=.25\linewidth,angle=0]{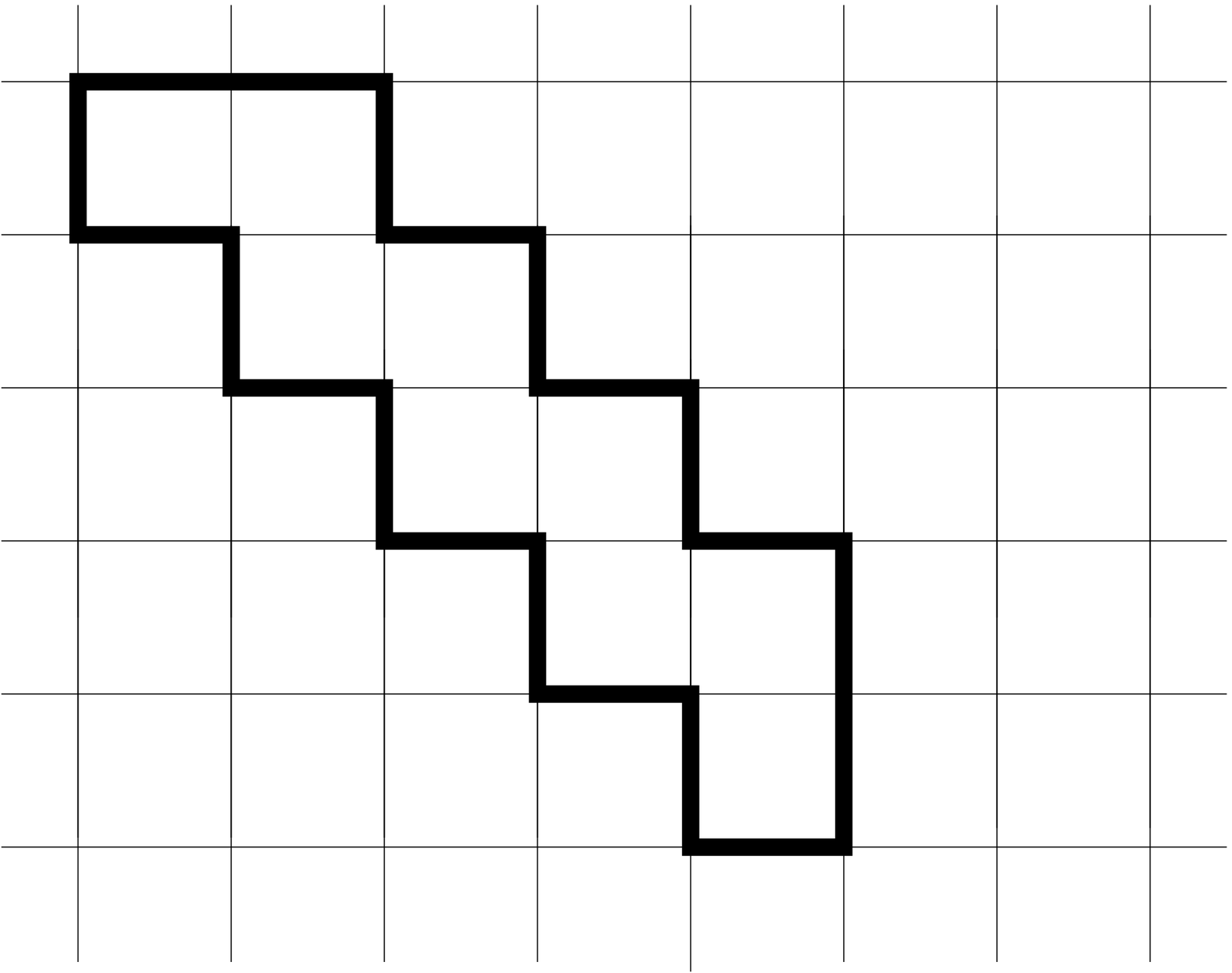}
\end{center}
  \caption{\emph{An illustration of the different Wilson loops that
we considered. Their contours are all polygonal, without 
multiple-intersections. They are squares, L-shapes 
(with legs of width $1$ lattice spacing and equal length, 
or lengths deviating by $1$ if the area $A$ is even), maximally 
anisotropic rectangles and stairs (again in two variants,
depending whether $A$ is even or odd).}}
\label{shapes}
\end{figure}

The Eguchi-Kawai equivalence to the model solved by Gross and 
Witten also implies the validity of the APD symmetry in the planar
limit. Again we testify if this symmetry can be observed
to set in (approximately) for the system sizes that we 
simulated. We consider four types of Wilson loops: \\
we denote them as {\em square loops, rectangular loops, 
stair loops} and {\em L-loops.} In particular we considered these
loops at the areas $A=4,$ $9,$ $16,$ $25$ $\dots$ in lattice units.
The rectangular loops 
are maximally anisotropic, i.e.\ their
shapes are rectangles of side lengths $1$ and $A$.\footnote{In 
Section 4 we will also consider rectangles with a fixed
ratio between the two side lengths.}
For the shapes of stair loops and L-loops we refer to Figure \ref{shapes}.
Note that these two loop types involve two slightly different
cases, depending on whether the area $A$ is even or odd.

All the four types of loops (with $A$ fixed) are related by APDs.
In particular the square and rectangles transform into each
other under $SL(2,R)$ on the plane.\\

\begin{figure}[ht!]
\hspace*{-2mm}
  \includegraphics[width=.36\linewidth,angle=270]{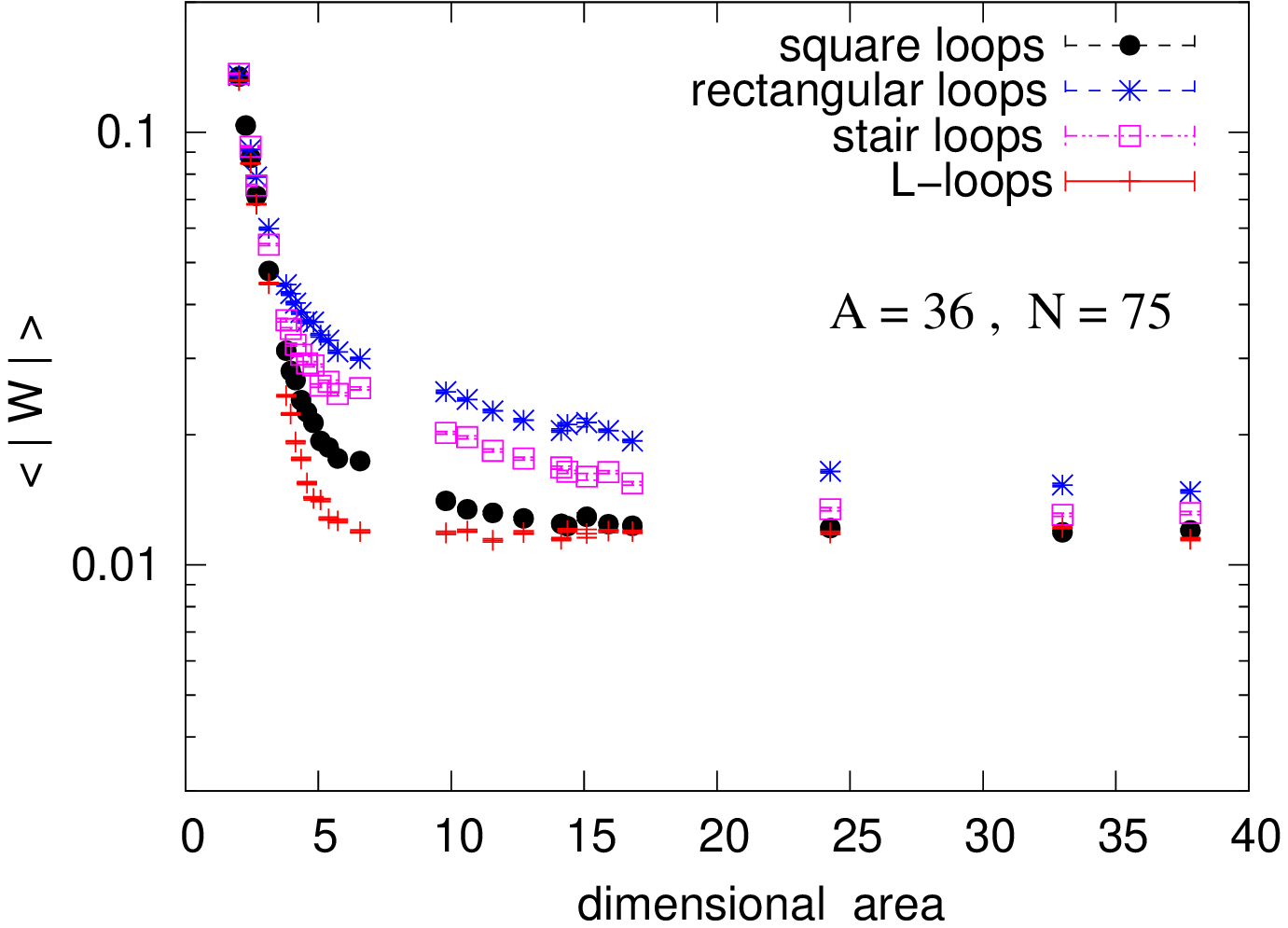} 
\hspace*{-2mm}
  \includegraphics[width=.36\linewidth,angle=270]{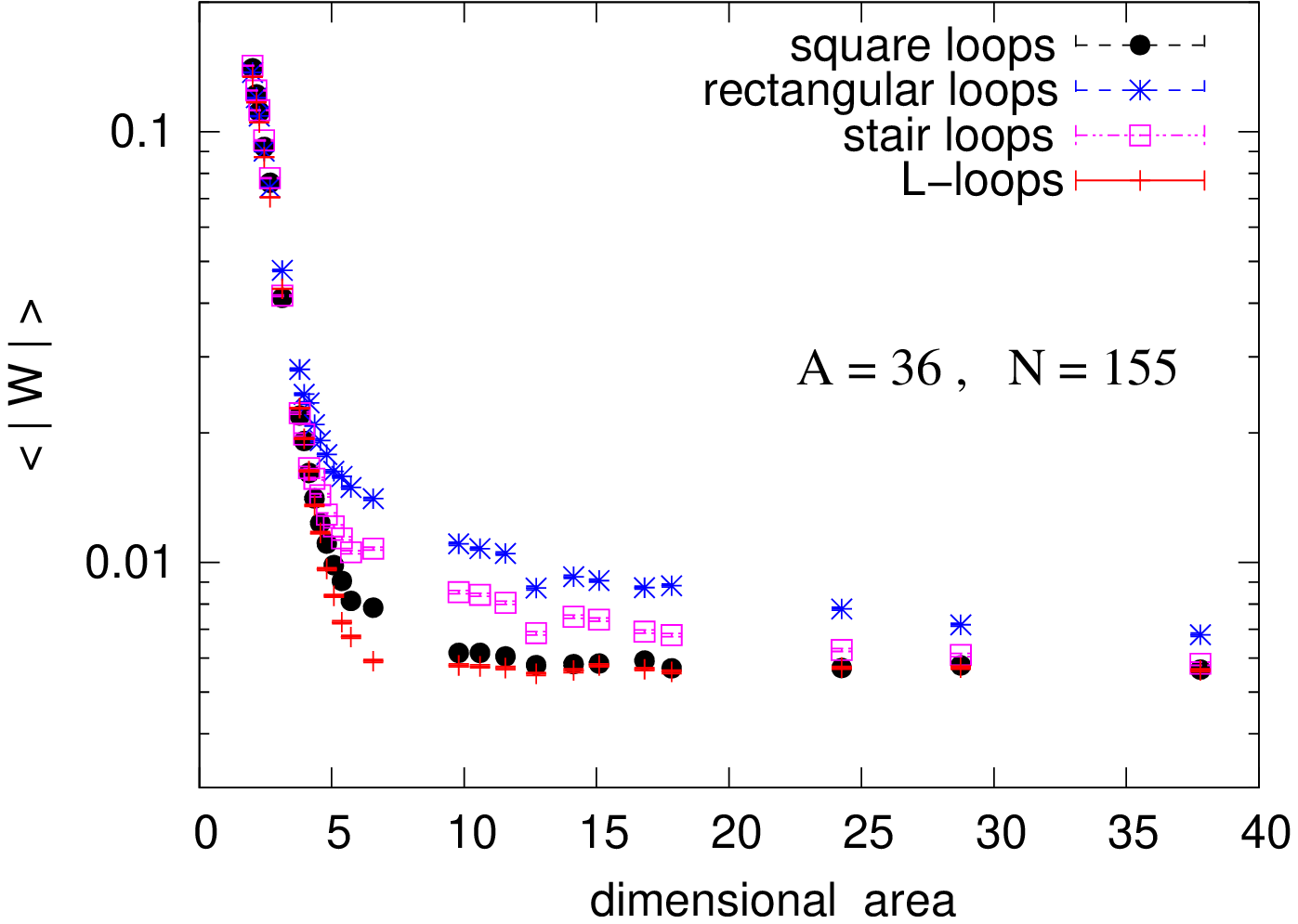}
\vspace*{-3mm}
  \caption{\emph{The absolute values of Wilson loops with different
shapes and lattice area $A=36$. They are plotted against the physical
area $36 a^{2}$, which is varied by using different values of $\beta$.
We show results for $N=75$ on the left, and for $N=155$ on the right.
The results for different shapes coincide best at small and at larger
areas, but they differ most around $36 a^{2} \approx 10$.}}
\label{figabs36}
\end{figure}

We now present numerical results for these Wilson loops $W$
as we approach the planar limit. As an example, we
first consider the absolute value $| W |$
as a function of the dimensional area
at a fixed lattice area of $A=36$.
Figure \ref{figabs36} shows results for $N = 75$ and $155$. 
We see that the absolute values
for the different shapes coincide for small and for large areas,
but they split apart at an intermediate physical
area of $ A a^{2} = {\cal O}(10)$. 

\begin{figure}[ht!]
\hspace*{-2mm}
  \includegraphics[width=.36\linewidth,angle=270]{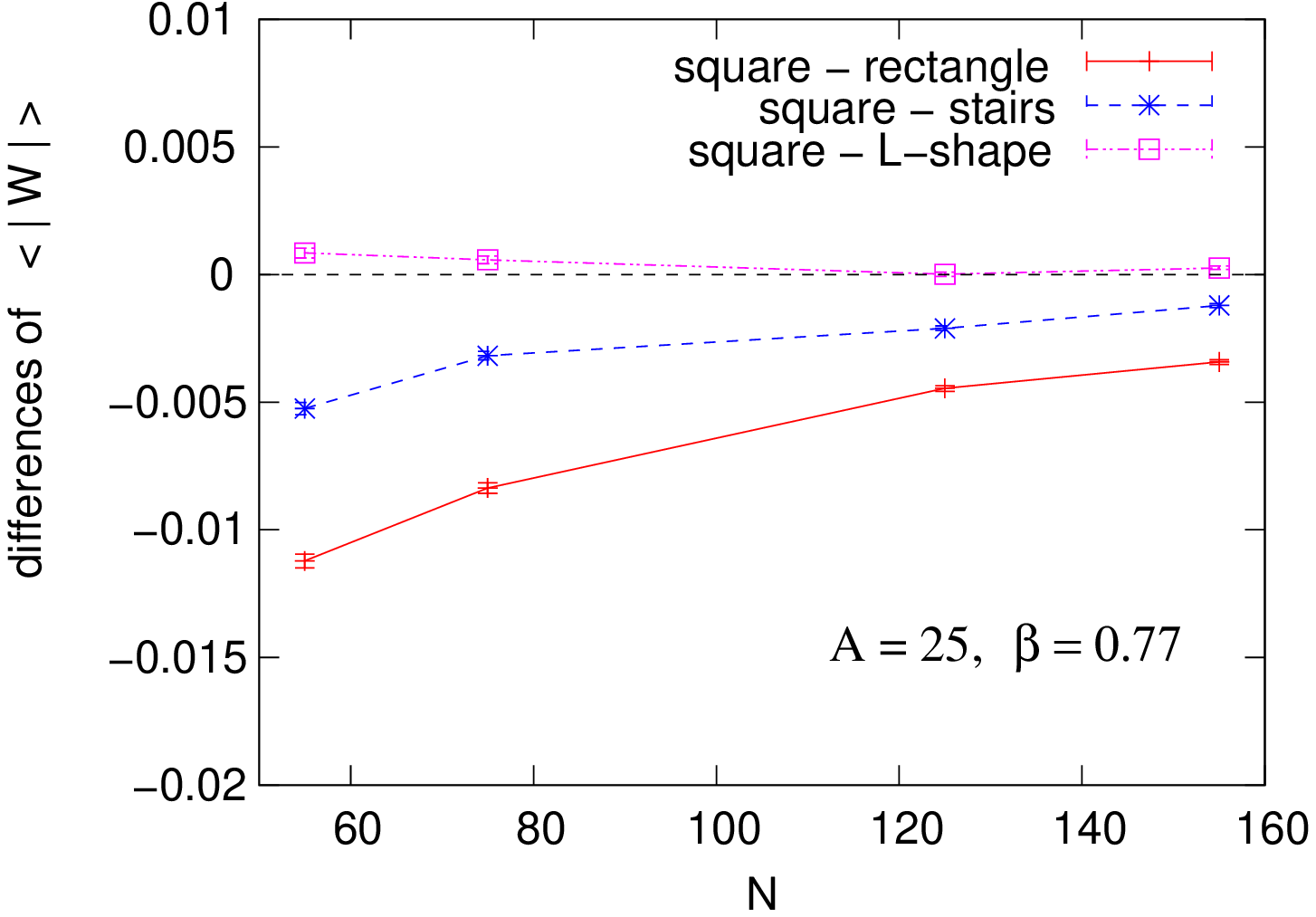}
\hspace*{-2mm}
  \includegraphics[width=.36\linewidth,angle=270]{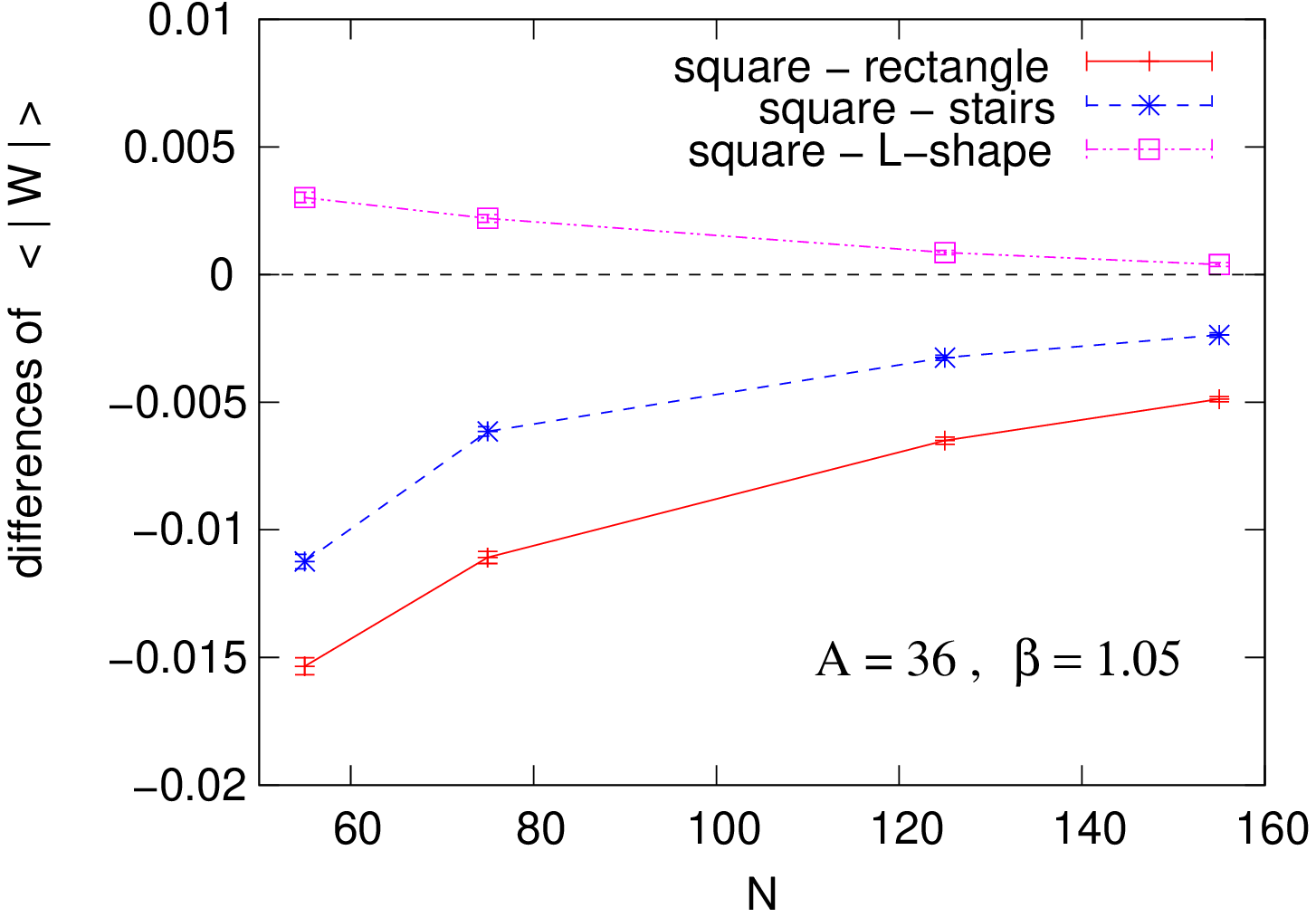}
\vspace*{-3mm}
  \caption{\emph{An illustration of the convergence towards
the APD symmetry in the planar limit. On the left
we show  the differences of $\, \langle |W| \rangle \, $ for 
Wilson loops with lattice area $A=25$ at $\beta = 0.77$, so the
dimensional area amounts to $Aa^{2} \simeq 9.81$. In the plot
on the right-hand-side the parameters are $A=36$ and
$\beta =1.05$, which implies almost the same area,
$Aa^{2} \simeq 9.79$. For increasing $N$ the differences
between the Wilson loops with different shapes decrease rapidly,
so we are approximating well the behaviour in the planar limit.}}
\label{planloopsabs}
\end{figure}

However, even in this intermediate regime the differences between the
Wilson loops at a fixed area converge to zero in the planar limit; 
examples for this behaviour are shown in Figure \ref{planloopsabs}.
Since $\beta $ was kept fixed in these plots, both the area in
lattice units and in dimensional units is constant. 

For the interpretation of our simulation results, we can therefore
rely on the scale (\ref{adim}) extracted from the planar limit. 
This allows us to proceed now to the investigation of the DSL, 
which describes the theory on a NC plane in the simultaneous UV 
and IR limit.


%

\section{The Double Scaling Limit}

We now employ the scale (\ref{adim}) provided by the Gross-Witten area law
(\ref{arealaw}); its use for the range of parameters under
consideration has been justified in Section 3.
This enables us to study the DSL (\ref{DSLeq}) to a continuous plane 
of infinite extent and finite non-commutativity. 

From Refs.\ \cite{NCQED2} we know the following properties
about the square loops:

\begin{itemize}

\item The observable $\langle W(I \times I) \rangle$ 
does indeed stabilise in the DSL. 
The existence of this universality class shows in particular
that the model is non-perturbatively renormalisable.

\item At small area, the absolute value 
$| \langle W ( I \times I) \rangle |$ follows an
area law. In that regime, which extends up to
$ (a I)^{2} \lsim 4$, the phase is practically zero.

\item For larger areas, $| \langle W (I\times I) \rangle |$
does not decay any further, but the phase starts to increase 
linearly in the area. It obeys the simple relation
\begin{equation}  \label{ABeq}
{\rm phase} = \frac{(aI)^{2}}{\theta} =  (aI)^{2} \cdot B \ ,
\end{equation}
where we symbolically introduced a magnetic field
$B = 1/ \theta$ across the plane. As we mentioned in Section 2,
this identification of the magnetic field has been 
implemented in string theory and in solid state physics.
The behaviour (\ref{ABeq}) just corresponds to the 
{\em Aharonov-Bohm effect}, which fits perfectly
into the magnetic interpretation of non-commutativity.
Ref.\ \cite{NCQED2} discovered this behaviour 
(unexpectedly) as a dynamical effect at low energy.

\end{itemize}

\begin{figure}[ht!]
\begin{center}
  \includegraphics[width=.55\linewidth,angle=270]{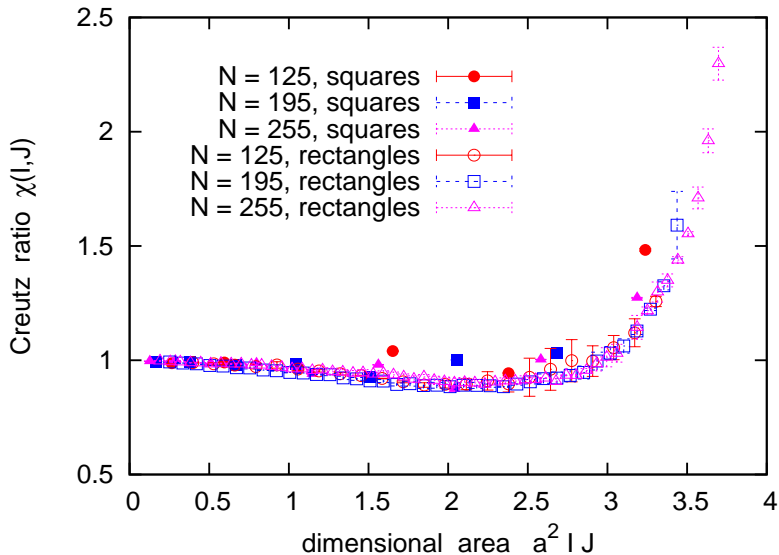}
\end{center}
\vspace*{-3mm}
\caption{\emph{The Creutz ratio $\chi (I,J)$ for Wilson loops with small 
areas, up to the transition at the end of the area law regime.
The parameter sets are $(N , \beta ) = (125, 3.91)$, $(195, 6.09)$
and $(255, 7.97)$, which all corresponds to $\theta \approx 2.6$,
so we are approaching a DSL. For both shapes, i.e.\ for $\chi (I,I)$
(squares) and for $\chi (2,J)$ (rectangles), we observe stable Creutz 
ratios in the DSL. At small area the string tension takes the same value
as in the planar limit. At moderate area the results from
square shaped and from extremely anisotropic rectangles begin to
differ a little.}}
\label{Creutz_phasefig}
\end{figure}

To revisit the second point --- the area law for small Wilson loops, 
where the phase is practically zero --- we measured the Creutz ratio 
\begin{equation}
\chi (I,J) = - \ln \Big[ \ \frac{ \langle W(I \times J) \rangle \ 
\langle W((I-1) \times (J-1)) \rangle }{\langle W((I-1) \times J)
\rangle \ \langle W(I \times (J-1)) \rangle } \ \Big] \ .
\end{equation}
This ratio singles out the string tension
$\sigma$ for decays $\propto \exp (- \sigma A)$, provided
that it is equivalent for the various rectangular Wilson loops
involved. 
Typical results for (nearly) square shaped Wilson loops, 
$\chi (I,I)$, as well as extremely anisotropic (rectangular) 
Wilson loops, $\chi (2,J)$, are shown in 
Figure \ref{Creutz_phasefig}.\footnote{We actually averaged
over $\chi(2,J)$ and $\chi(J,2)$ in order to increase the statistics.}
For both shapes we find a stable behaviour as we increase $N$ at fixed
$\theta$, which suggests that our results can safely be extrapolated
to the DSL. Deeply inside the area law regime
we obtain $\sigma \simeq 1$. Hence in this range the behaviour in the 
DSL coincides with the planar limit. We observe, however,
a marked deviation from it as the area approaches
the transition to the regime of the Aharonov-Bohm type behaviour.

At moderate area the Creutz ratios for the squares
and the rectangles differ a little. 
This is a first observation hinting at shape independence for 
rectangles deeply inside the area law regime, but not beyond.

Regarding the behaviour at large areas, which is very specific
to the NC plane,
one may wonder why the short-ranged non-commutativity
has striking effects on the large rather than the small Wilson loops.
UV/IR mixing \cite{UVIR} is apparently at work, even though
the perturbative expansion of this model can be formulated
without divergences.
This suggests that UV/IR mixing occurs non-perturbatively, and it belongs
therefore to the fundamental nature of NC field theory.
This is in agreement with analytic \cite{GuSo} and numerical 
\cite{NCphi4,Antonio} results for the NC $\lambda \phi^{4}$ model,
and for 4d NC QED \cite{NCQED4}.\\

\begin{figure}[htbp!]
\begin{center}
\vspace*{2mm}
\includegraphics[width=.55\linewidth,angle=270]{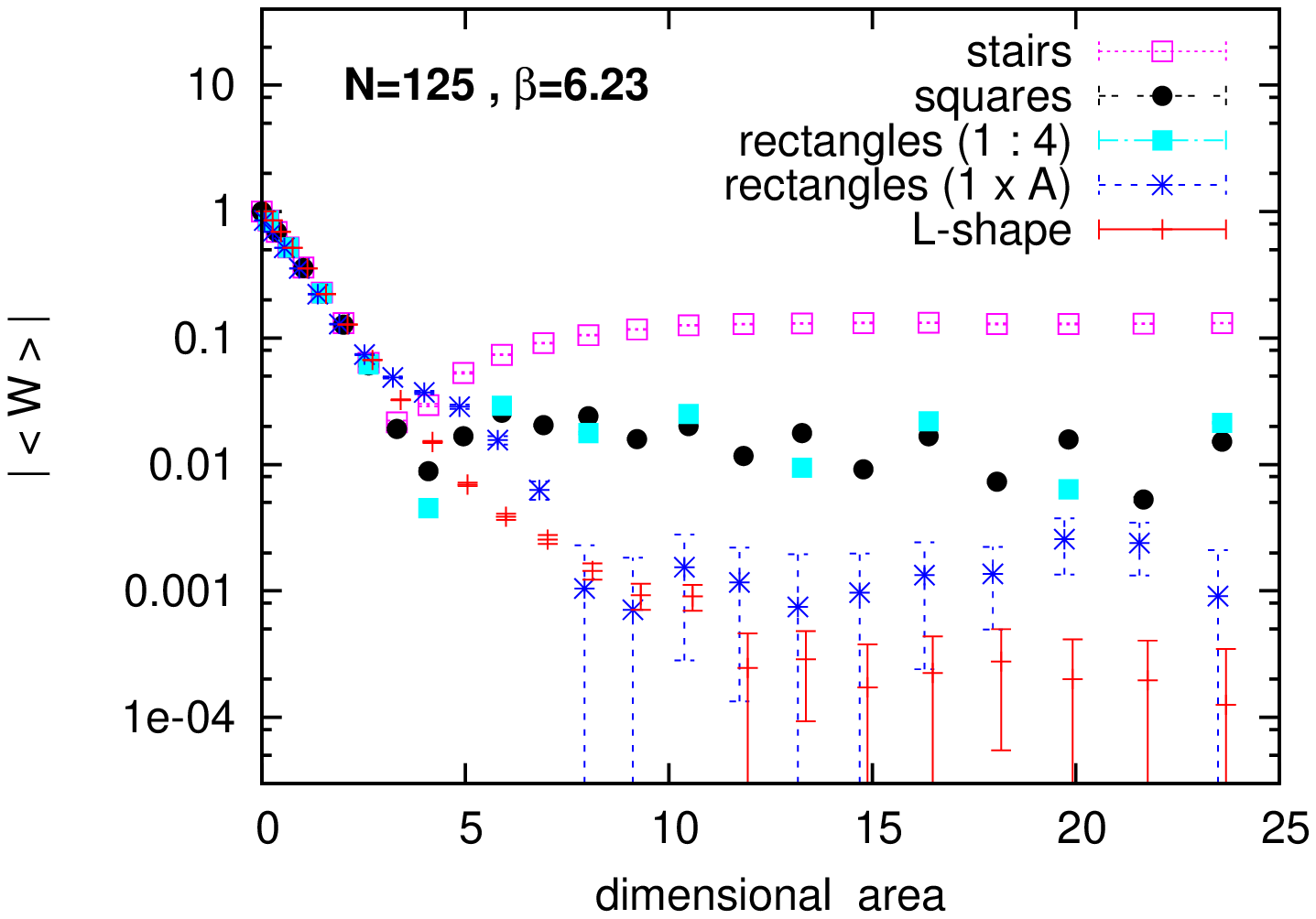}
\vspace*{2mm} \\
\includegraphics[width=.55\linewidth,angle=270]{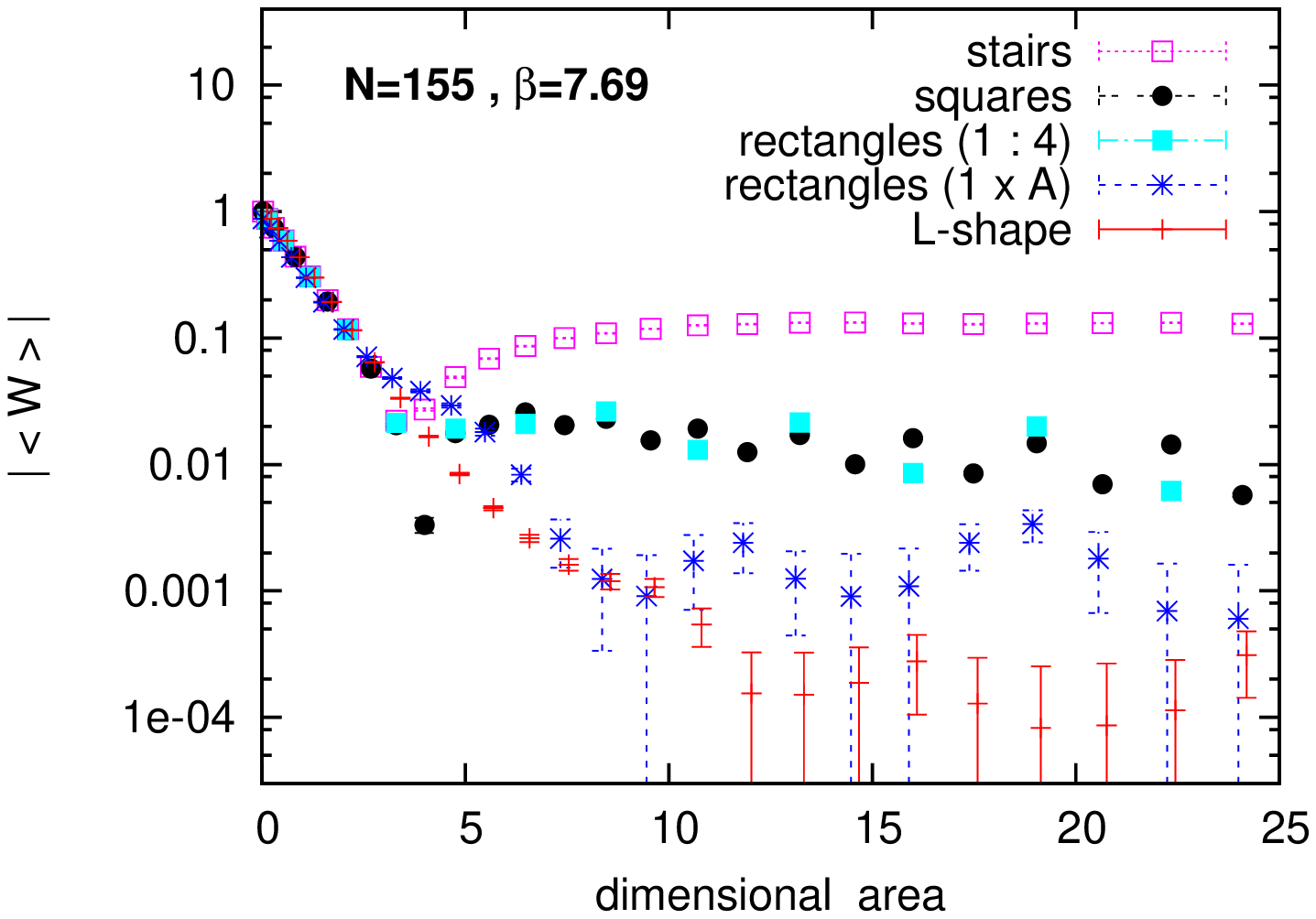}
\end{center}
\caption{\emph{The absolute values of various Wilson loops
at a fixed non-commutativity parameter $\theta = 1.63$.
On top we show results at $N=125$ and below at $N=155$. 
As we increase $N$ the dimensional volume grows and the lattice spacing
shrinks, so that we approach simultaneously the limits to the
continuum and to infinite volume (UV and IR limit).
The striking similarity of these plots confirms that
DSL convergence is reached. The results show a clear shape dependence 
beyond the area law regime, and therefore the breaking of APD symmetry.}}
\label{DSLabstheta1}
\end{figure}

\begin{figure}[ht!]
\begin{center}
\vspace*{2mm}
\includegraphics[width=.55\linewidth,angle=270]{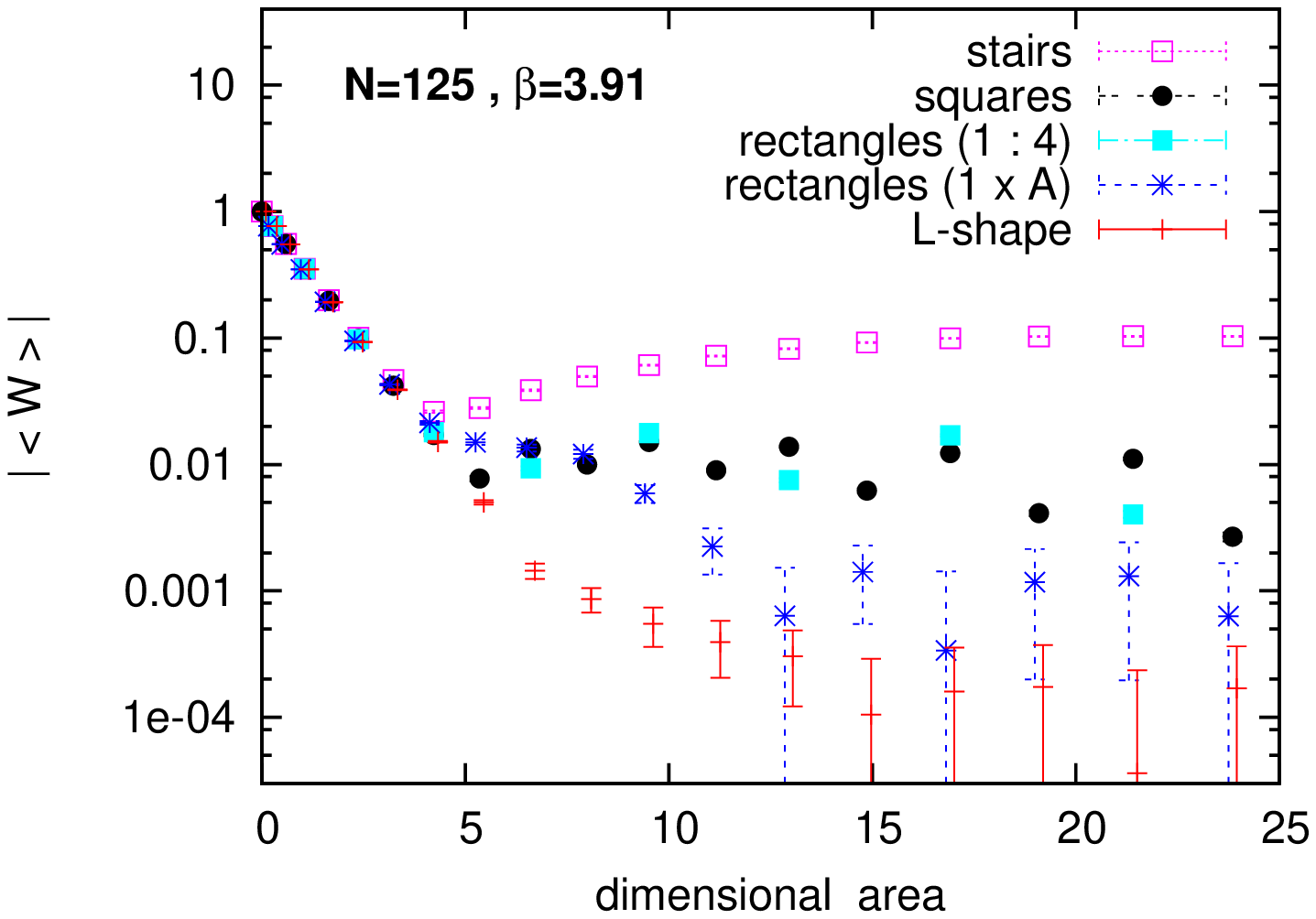}
\vspace*{2mm}
\includegraphics[width=.55\linewidth,angle=270]{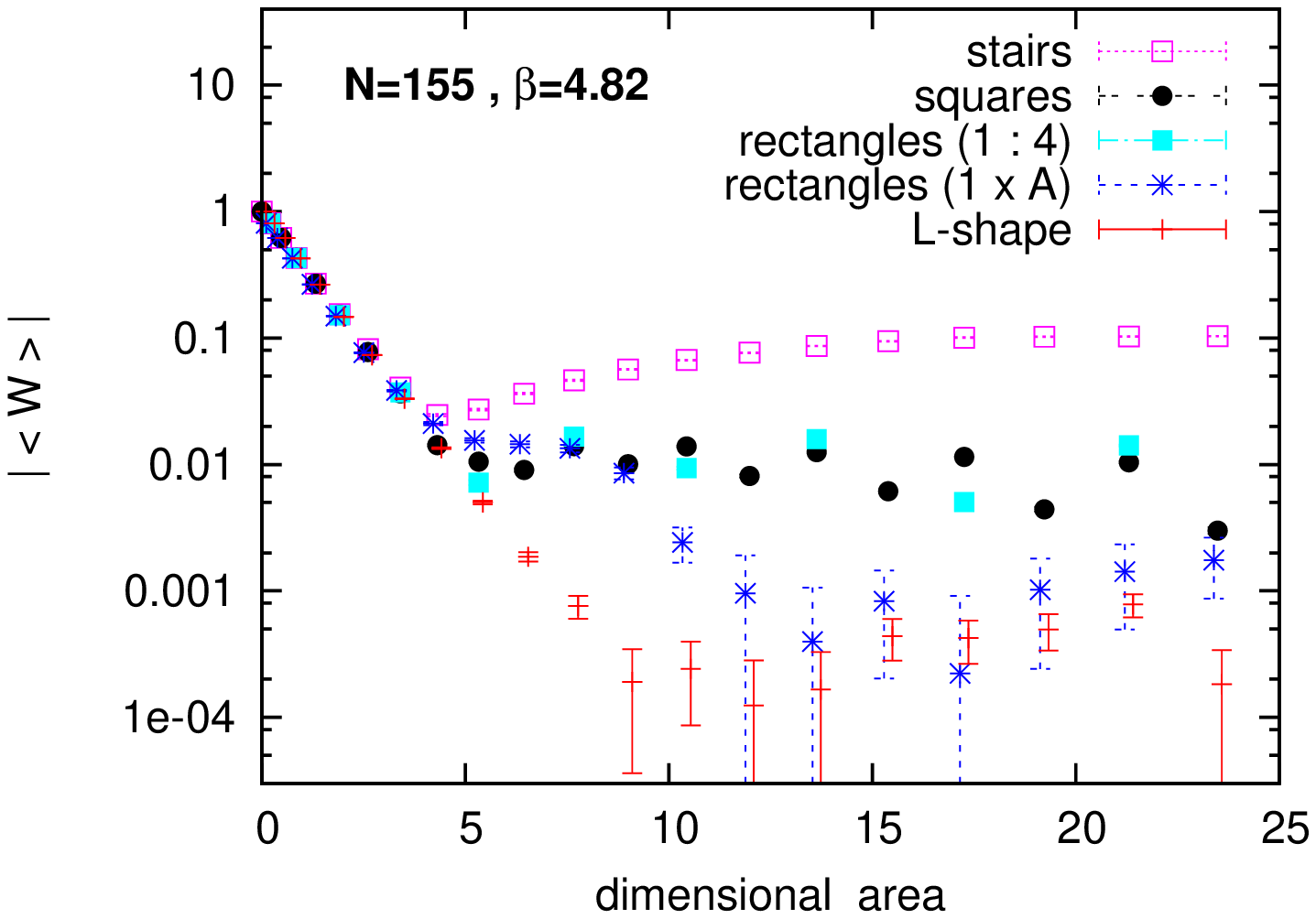}
\end{center}
\caption{\emph{The analogous plots to Figure \ref{DSLabstheta1},
but now at $\theta = 2.63$.
Again the results clearly confirm a shape dependence beyond the
area law regime, which is stable as we increase $N$ towards the DSL.}}
\label{DSLabstheta2}
\end{figure}

\begin{figure}[ht!]
\begin{center}
\vspace*{2mm}
\includegraphics[width=.55\linewidth,angle=270]{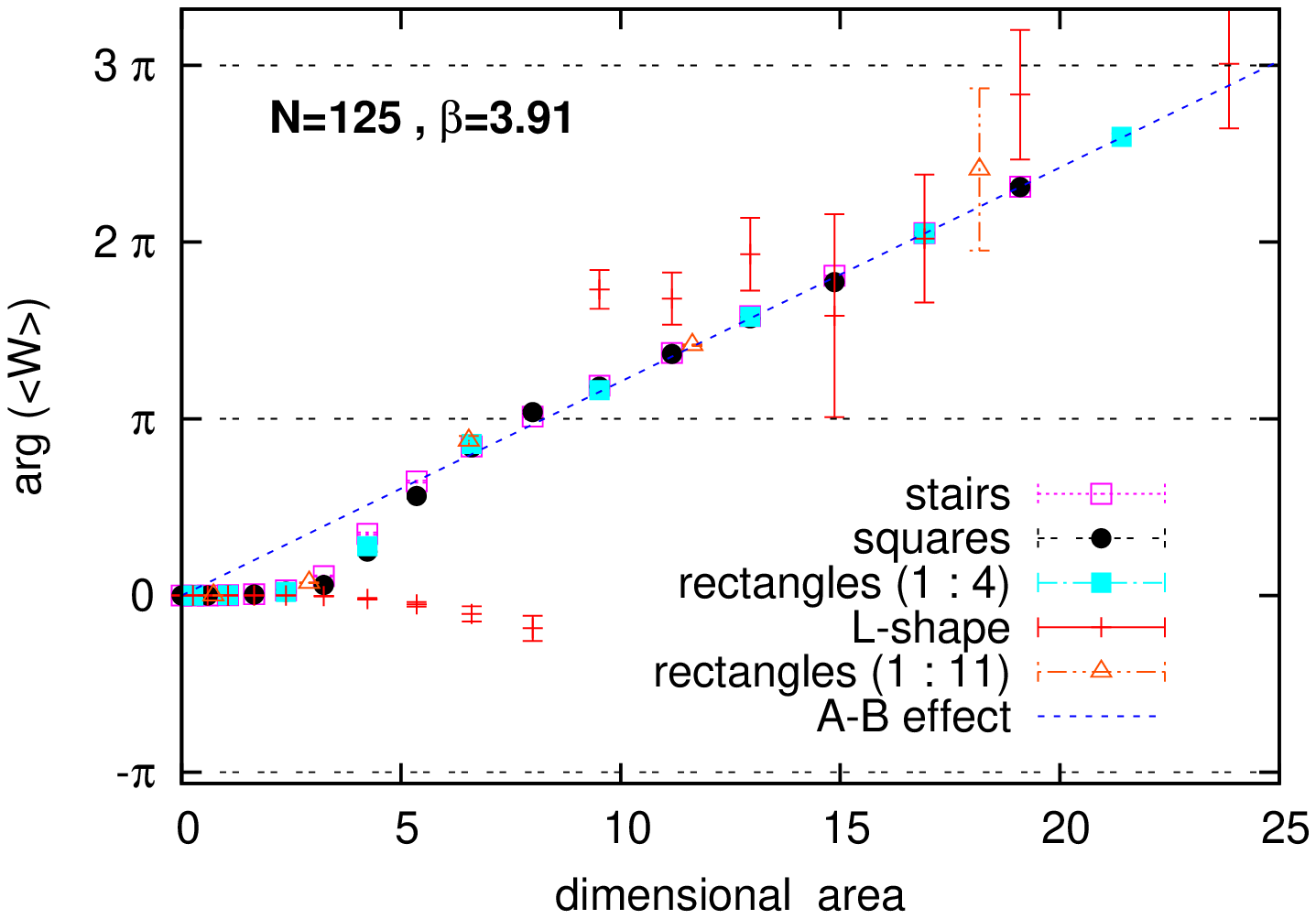}
\vspace*{2mm} \\
\includegraphics[width=.55\linewidth,angle=270]{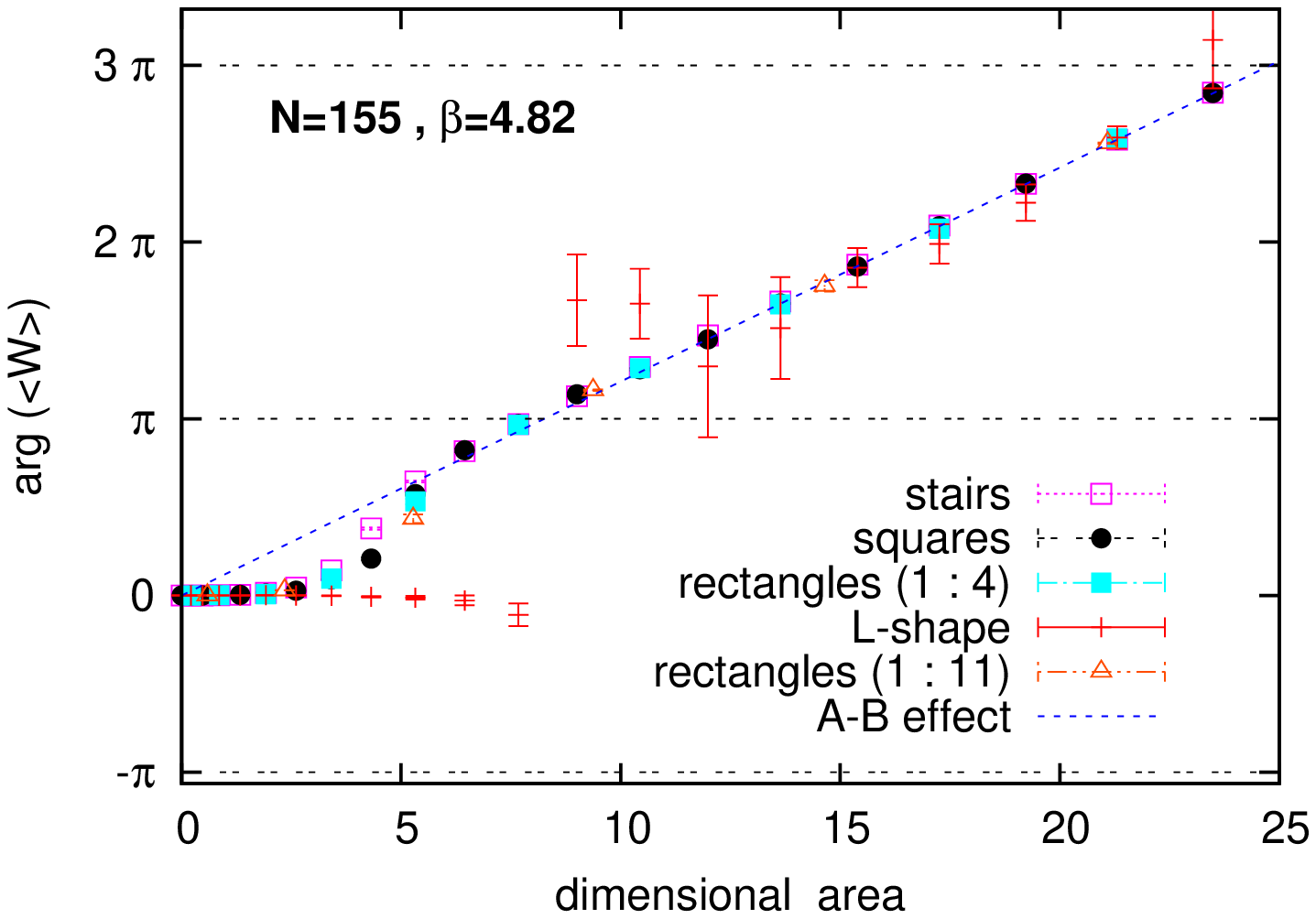}
\end{center}
\caption{\emph{The phases of various Wilson loops
at $\theta = 2.63$ for $N=125$ (on top) and $N=155$ (below).
The Aharonov-Bohm type behaviour in eq.\ (\ref{ABeq}) at large area 
is very well confirmed for the shapes shown here. 
They include results for rectangular loops with a fixed ratio of $4$ 
and a ratio of $11$ between the side lengths, 
so that its shape is invariant on the way
to the DSL. The L-shape, on the other hand, becomes infinitesimally
thin in this limit. Its phase follows eq.\ (\ref{ABeq}) with some
fluctuations.}}
\label{DSLphases}
\end{figure}

At last we proceed to a systematic study of the fate of the APD
symmetry in the DSL. Figures \ref{DSLabstheta1} and \ref{DSLabstheta2}
show results for the absolute values of Wilson loop expectation values,
$\, | \langle W \rangle | \, $, with respect to the shapes illustrated 
in Figure \ref{shapes}. Figure \ref{DSLabstheta1} refers to
a fixed non-commutativity parameter $\theta = 1.63$, and it shows
results for two values of $N$. For increasing $N$ the dimensional
volume is enlarged and the lattice becomes finer (because
$N a^{2}$ is kept constant), so we approach the DSL. 
The corresponding results at $\theta =2.63$ are presented
in Figure \ref{DSLabstheta2}. For the latter $\theta$ value we show 
in addition the Wilson loop phases $\, {\rm arg} ( \langle W \rangle ) \,$
in Figure \ref{DSLphases}. These figures demonstrate that
the Wilson loops for $N=125$ and $N=155$ --- plotted against the 
dimensional area --- are almost identical. 
This shows that we are indeed in the asymptotic regime of the 
DSL. We can therefore be confident that our
results reveal the behaviour in the continuous NC plane.
This confidence will be further substantiated by results for
variants of these observables to be presented in the Section 5.

As soon as the area exceeds the area law regime, we observe a 
clear distinction between the absolute values $| \langle W \rangle |$
for different shapes, see Figures \ref{DSLabstheta1} and 
\ref{DSLabstheta2}.\footnote{We note that the total area of the
system amounts to $V = \pi N \theta$, hence for our parameters
$V \gg \theta$ is granted.}
This distinction occurs for any pair of the contour types
considered. We show again
the rectangles of the form $1 \times A$ (in lattice units),
which become infinitesimally narrow in the DSL (in dimensional units).
In addition we also include rectangles with a fixed ratio of $4$ between
the side lengths, which keep an invariant shape as we approach the DSL.
The different expectation values for the 
narrow rectangles with the same area show clearly
that the APD symmetry breaks.

The distinction between squares and rectangles of fixed side ratio
in these plots is less striking than the other cases. However,
exactly these rectangles are directly relevant to explore
the fate of the symmetry subgroup $SL(2,R)$, hence we will focus
on them specifically in Section 6.

Next we discuss the phases $\, {\rm arg} (\langle W \rangle ) \, $, 
which we show for $\theta = 2.63$ in Figure \ref{DSLphases}.
As the area increases beyond the area law regime,
the square shapes, rectangles with fixed side ratio and the stairs 
follow very well the Aharonov-Bohm type behaviour corresponding 
to eq.\ (\ref{ABeq}),  which had been observed earlier for squares 
and certain rectangles \cite{NCQED2}. The L-shape does not agree
optimally, but its behaviour is reasonably close, in particular
on the finer lattice which corresponds to $N=155$.
We remark that the extremely anisotropic rectangles considered
earlier lead to strong deviations from eq.\ (\ref{ABeq}), as we
are going to show in the Section 5 (Figure \ref{rectphase}).
Apparently shapes which become extremely thin (in physical units) 
as we approach the DSL can lead to such features (although this 
is not the case for the stair loops).

In any case, our results for the phases are very similar for the 
different $N$ values, 
so they confirm that we are in an asymptotic window of the DSL.
The phase of the Wilson loop has a much stronger trend
towards (at least partial) APD symmetry than the absolute value.
As far as we could check, the phase (alone) is well
compatible with this symmetry for fixed shapes in dimensional
units, with a finite extent in each direction.
But of course the differences in $\, | \langle W \rangle | \,$
are sufficient to discard APD symmetry
as a basic property of this theory.

\section{APD symmetry breaking for alternative observables}

Our results for the standard observables $\, | \langle W \rangle | \,$
and $\, {\rm arg} ( \langle W \rangle ) \,$
were presented in Section 4 and illustrated
the shape dependence of Wilson loops with the same area (in
dimensional units). We gave evidence for this effect to persist
in the DSL. 

In this section we present the corresponding results for
$ \langle |W| \rangle $ and $\langle {\rm arg} (W) \rangle$, which are
somewhat different observables. The quantities of Section 4
are tractable in perturbation theory, and these are therefore
the observables that have been addressed in Refs.\ 
\cite{bnt1,bnt2,adm,noi,CGSS,RicSza,ADM2}. Moreover they were
measured in the previous numerical study in 
Ref.\ \cite{NCQED2}. On the other hand, the quantities
of this section are generally not considered in analytic work.
In the appendix we comment on the prospects of the perturbative
treatment of $ \langle |W| \rangle $.
However, $ \langle |W| \rangle $ and $\langle {\rm arg} (W) \rangle$  
can be handled numerically without specific problems,
and they also represent valid physical 
observables.\footnote{Actually we considered the observables
of this section already in Section 3 when we discussed the validity
of the scale identified from the planar limit (Figures \ref{figabs36} 
and \ref{planloopsabs}).}
For instance, in lattice gauge theory it is also 
usual to measure $\langle |P| \rangle$ ($P$ being the Polyakov
loop) --- its magnitude serves as a criterion to distinguish the
phases of confinement and deconfinement.

We add these observables here in order to supply further strength
to our observation of APD symmetry breaking. The new
observables are suitable for this purpose, in particular 
because $\langle |W| \rangle $ does not become as tiny as
$| \langle W \rangle |$ at moderate and large area, which leads
to smaller relative errors.
 
\begin{figure}[ht!]
\begin{center}
\vspace*{2mm}
\includegraphics[width=.51\linewidth,angle=270]{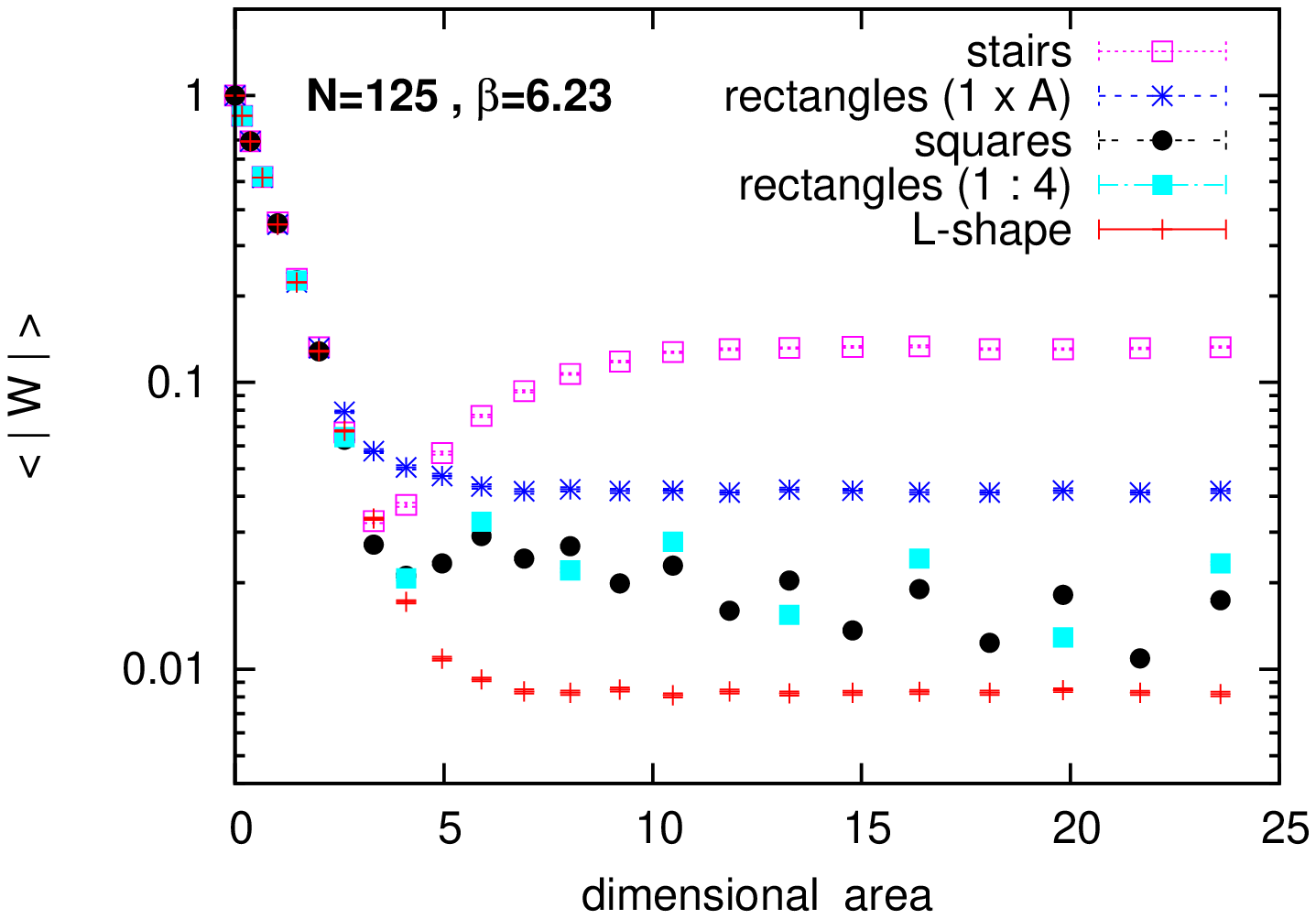}
\vspace*{2mm} \\
\includegraphics[width=.51\linewidth,angle=270]{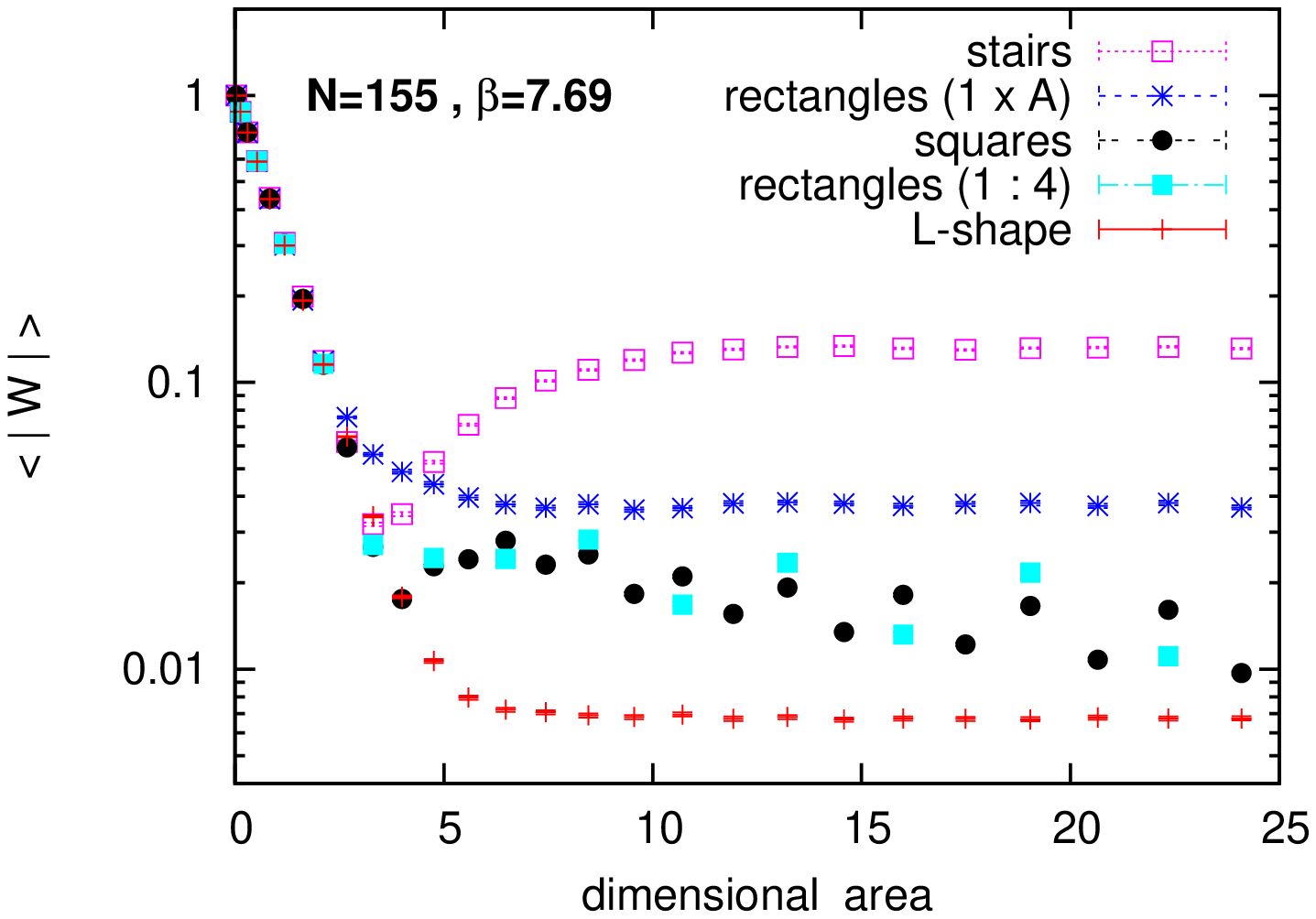}
\end{center}
\caption{\emph{The expectation values $\, \langle |W| \rangle \,$
for various Wilson loops at a fixed non-commutativity parameter 
$\theta = 1.63$. On top we show results at $N=125$ and below at
$N=155$. The striking similarity of these plots confirms that
DSL convergence is reached. The results reveal a clear shape dependence 
beyond the area law regime also for this observable.}}
\label{DSLabstheta1A}
\end{figure}

\begin{figure}[ht!]
\begin{center}
\vspace*{2mm}
\includegraphics[width=.51\linewidth,angle=270]{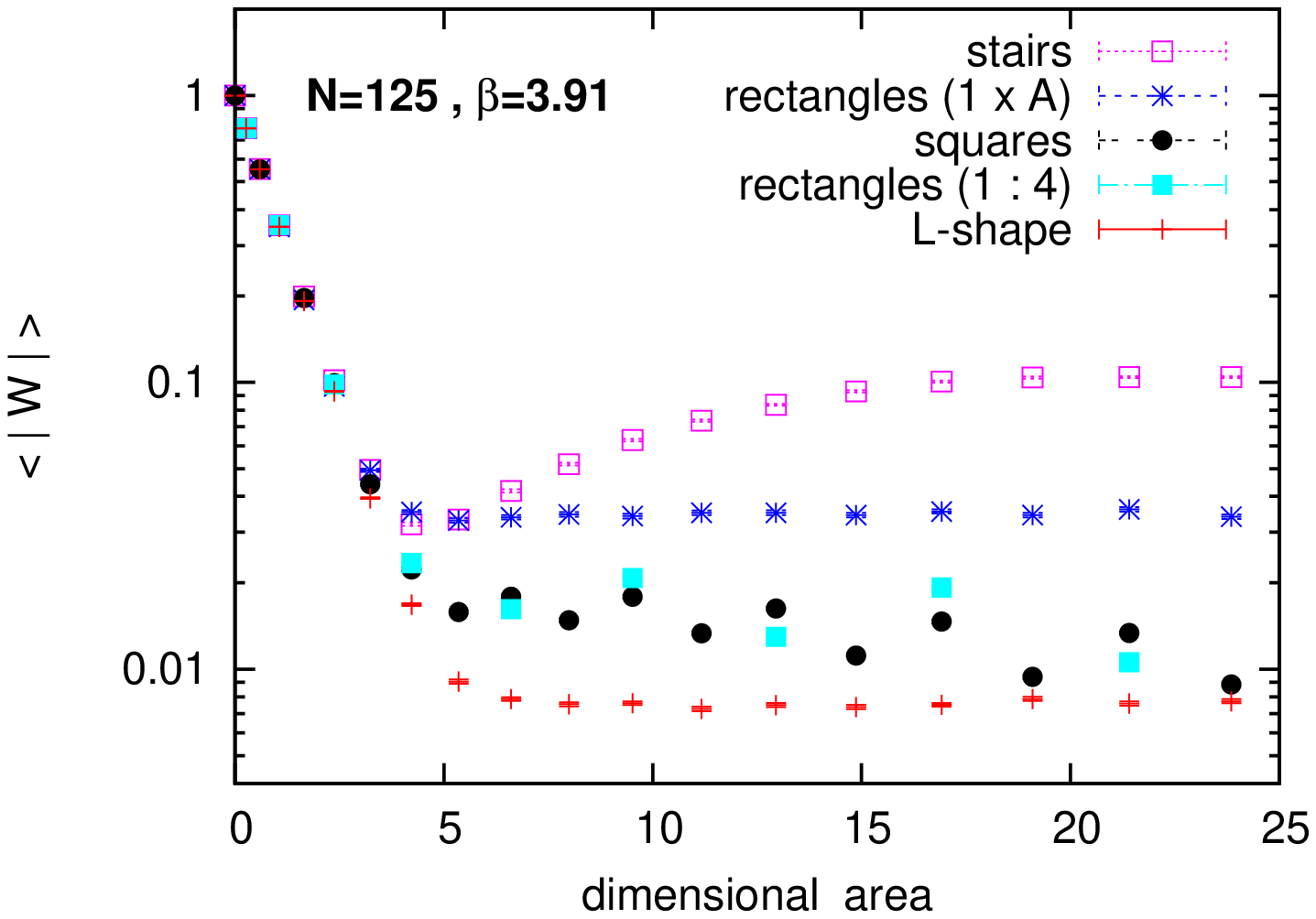}
\vspace*{2mm} 
\includegraphics[width=.51\linewidth,angle=270]{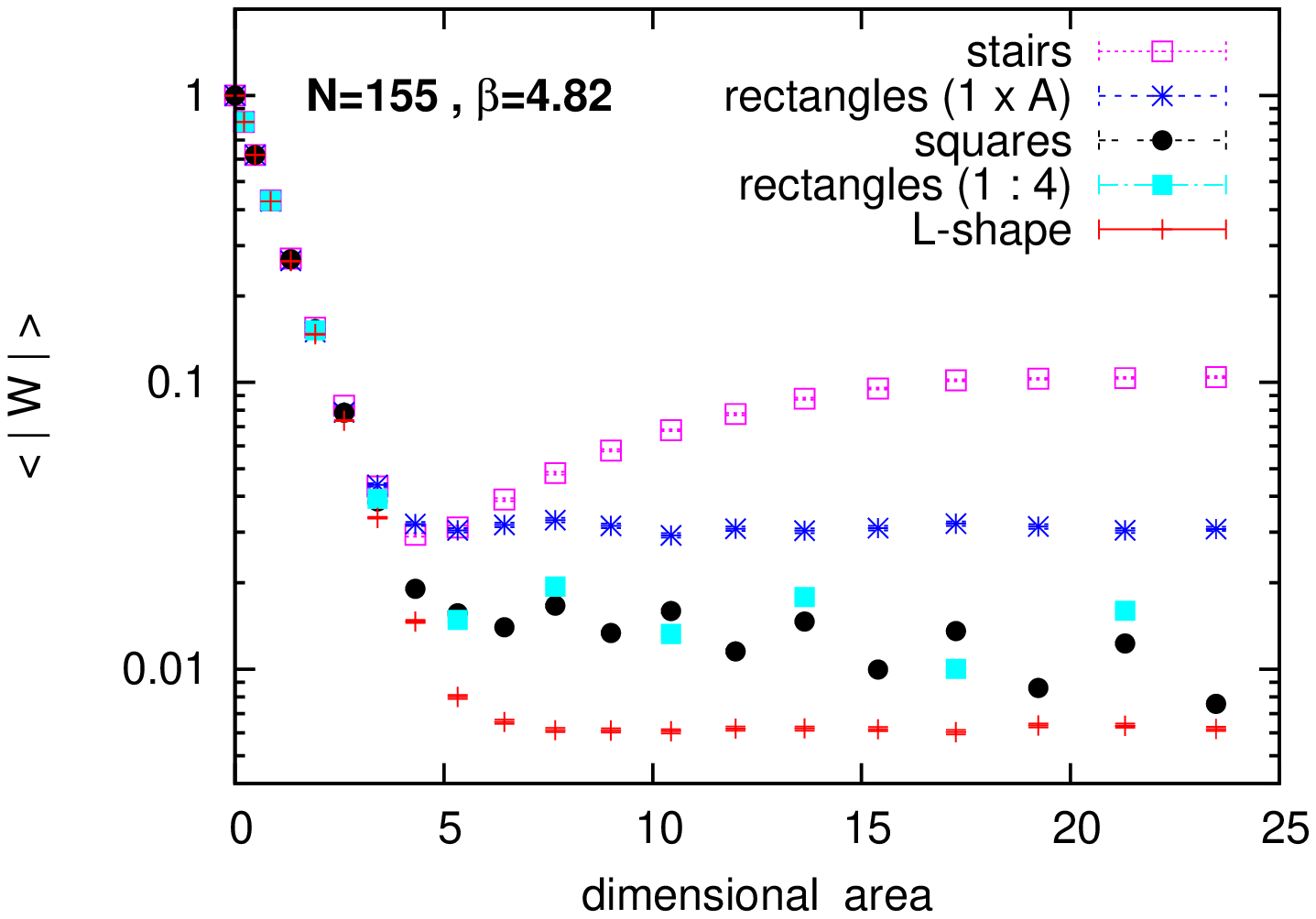}
\end{center}
\caption{\emph{The analogous plots to Figure \ref{DSLabstheta1A},
but now at $\theta = 2.63$.
The results affirm once more a shape dependence beyond the
area law regime, which is stable as we increase $N$.
In particular the differences at fixed area do not shrink as we
increase $N$ towards the DSL, in contrast to the planar limit
behaviour in Figure \ref{planloopsabs}.}}
\label{DSLabstheta2A}
\end{figure}

We first present the results for $\langle |W| \rangle $ for 
two values of $N$ at $\theta =1.63$ (Figure \ref{DSLabstheta1A})
and at $\theta =2.63$ (Figure \ref{DSLabstheta2A}), in analogy to
Figures \ref{DSLabstheta1} and \ref{DSLabstheta2}.
The new observables are practically identical to those of Section
4 at small area, where the phase is tiny throughout the
Monte Carlo history. They differ, however, as the area grows beyond 
this regime. There $\langle |W| \rangle $ is significantly
larger, which enables a better distinction (without overlapping
error bars). But the qualitative behaviour is very similar to Section 4.

\begin{figure}[ht!]
\begin{center}
\vspace*{2mm}
\includegraphics[width=.51\linewidth,angle=270]{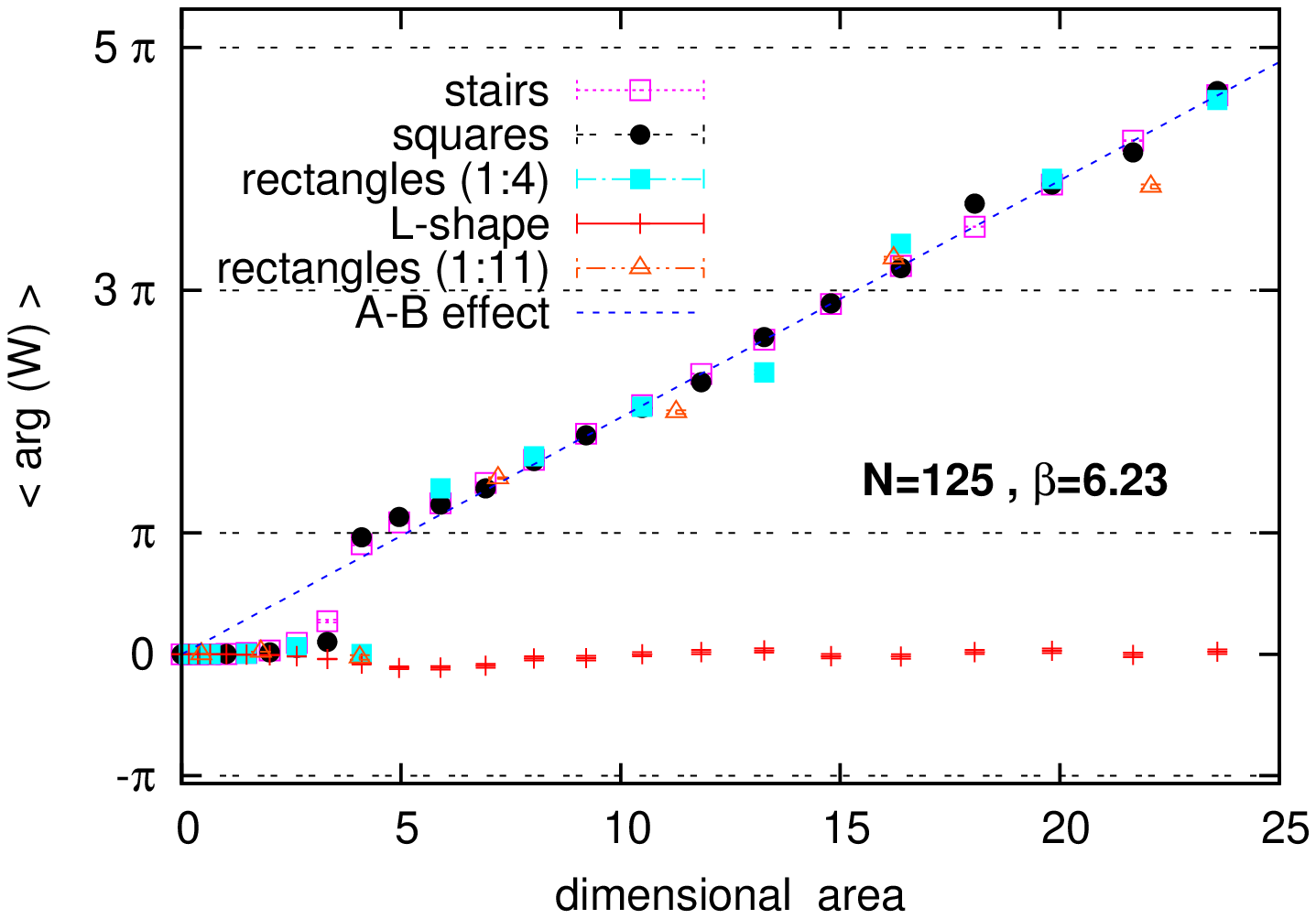}
\vspace*{2mm} \\
\includegraphics[width=.51\linewidth,angle=270]{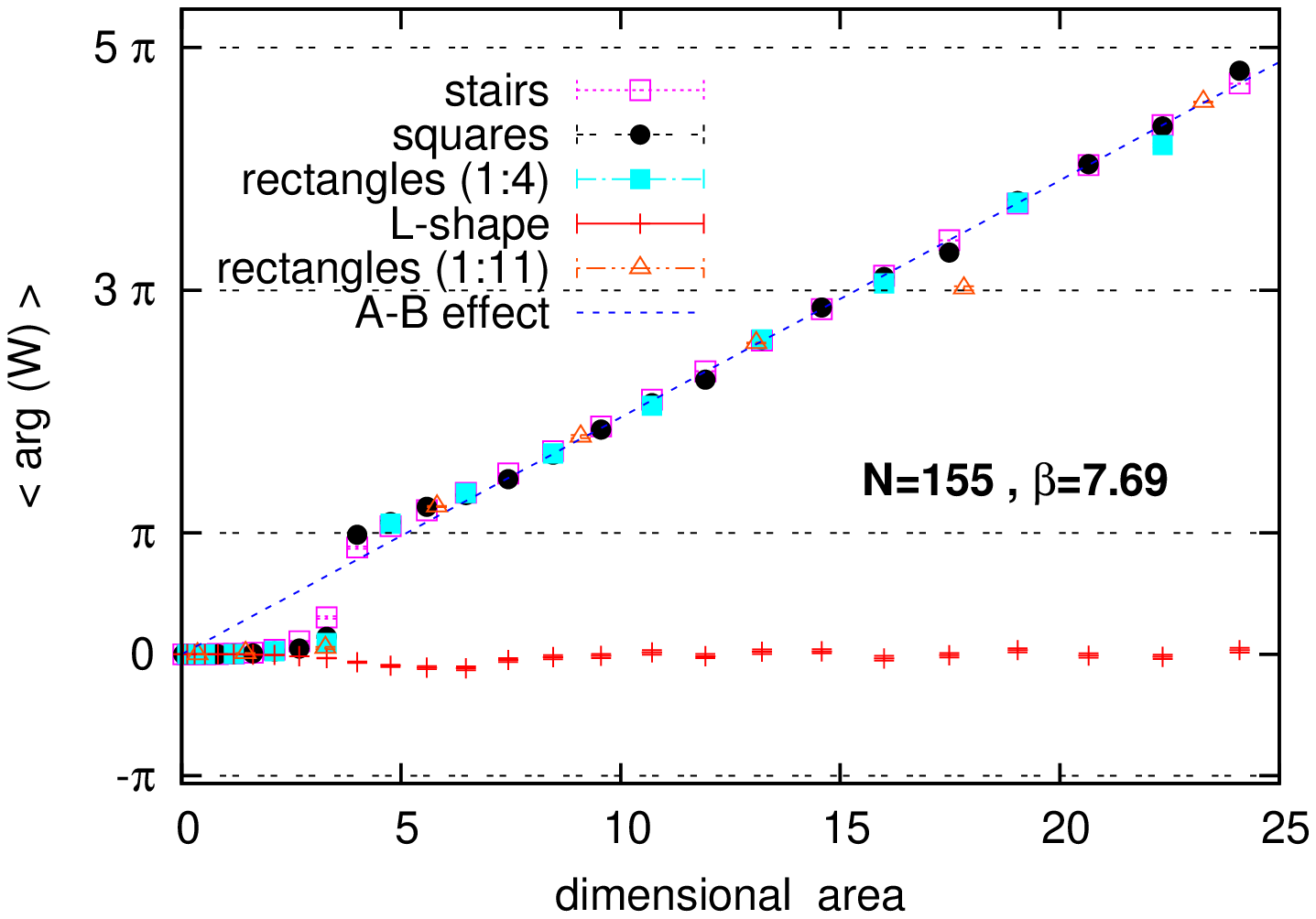}
\end{center}
\caption{\emph{The phases $\langle {\rm arg} (W) \rangle$
of various Wilson loops
at $\theta = 1.63$ for $N=125$ (on top) and $N=155$ (below).
The Aharonov-Bohm type behaviour in eq.\ (\ref{ABeq}) at large area 
holds well also for this phase for the stair loops,
the squares and the rectangular loops with a fixed side ratios of 
$4$ and of $11$.
On the other hand, the average phase of the L-shape loops remains
very small even at large areas.}}
\label{DSLphasesA}
\end{figure}

Figure \ref{DSLphasesA} shows the phases $\, \langle {\rm arg} (W) 
\rangle \,$ at the same parameters as in Figure \ref{DSLabstheta1A}. 
It can be compared to the phase ${\rm arg} (\langle W \rangle )$ 
presented before in Figure \ref{DSLphases} (for $\theta = 2.63$). 
Also this phase follows closely the
Aharonov-Bohm type behaviour of eq.\ (\ref{ABeq}) for the squares,
the stairs loops and for the rectangle with a fixed side ratios.
(For the squares this behaviour was also observed
in 4d NC $U(1)$ gauge theory \cite{NCQED4}.)
Again the phase of the L-shape loops deviates. It fluctuates strongly 
in the Monte Carlo history all over the interval $( - \pi , \pi ]$ ,
keeping its mean values close to zero.

\begin{figure}[ht!]
\hspace*{-2mm}
  \includegraphics[width=.36\linewidth,angle=270]{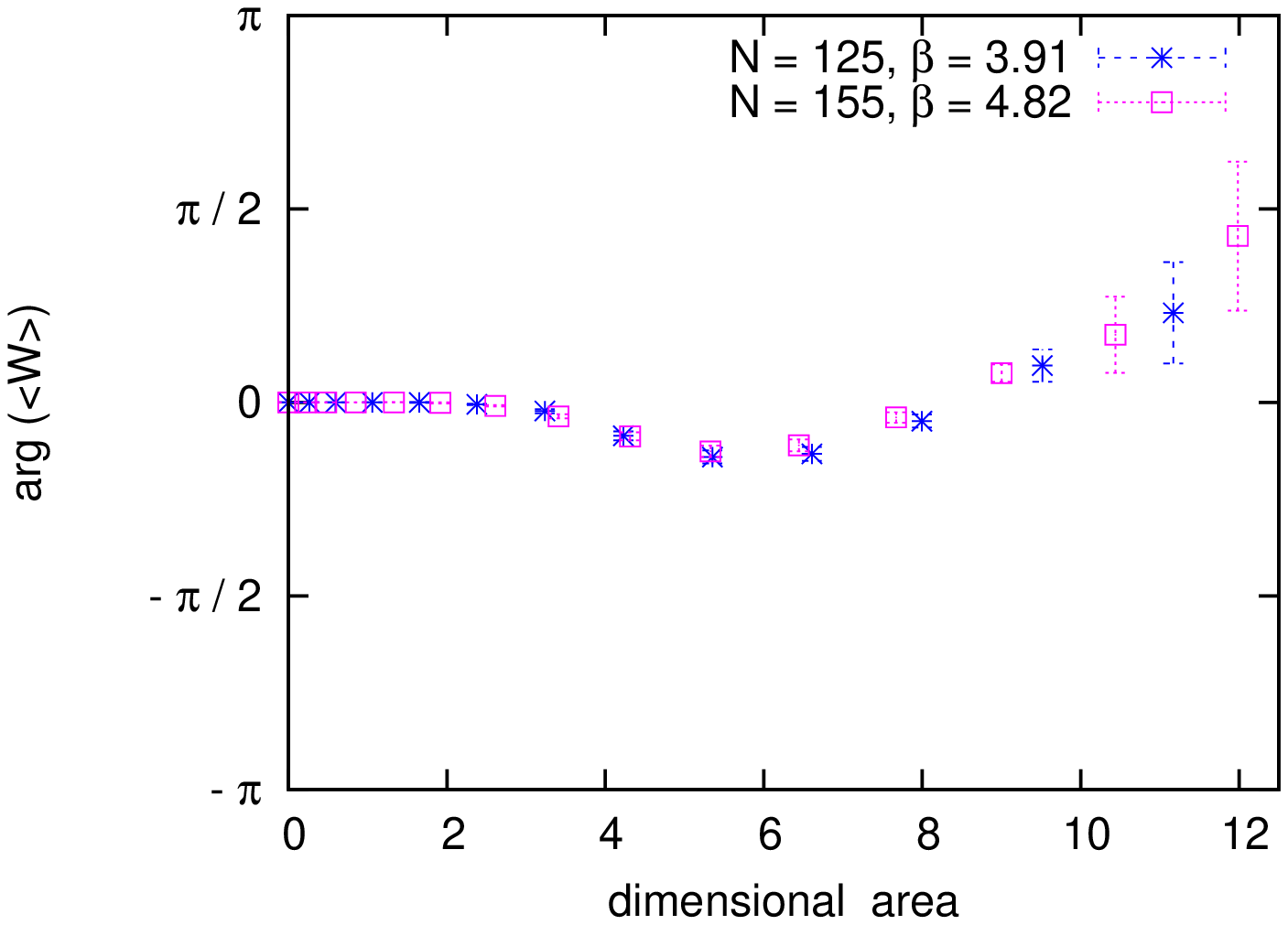} 
\hspace*{-2mm}
  \includegraphics[width=.36\linewidth,angle=270]{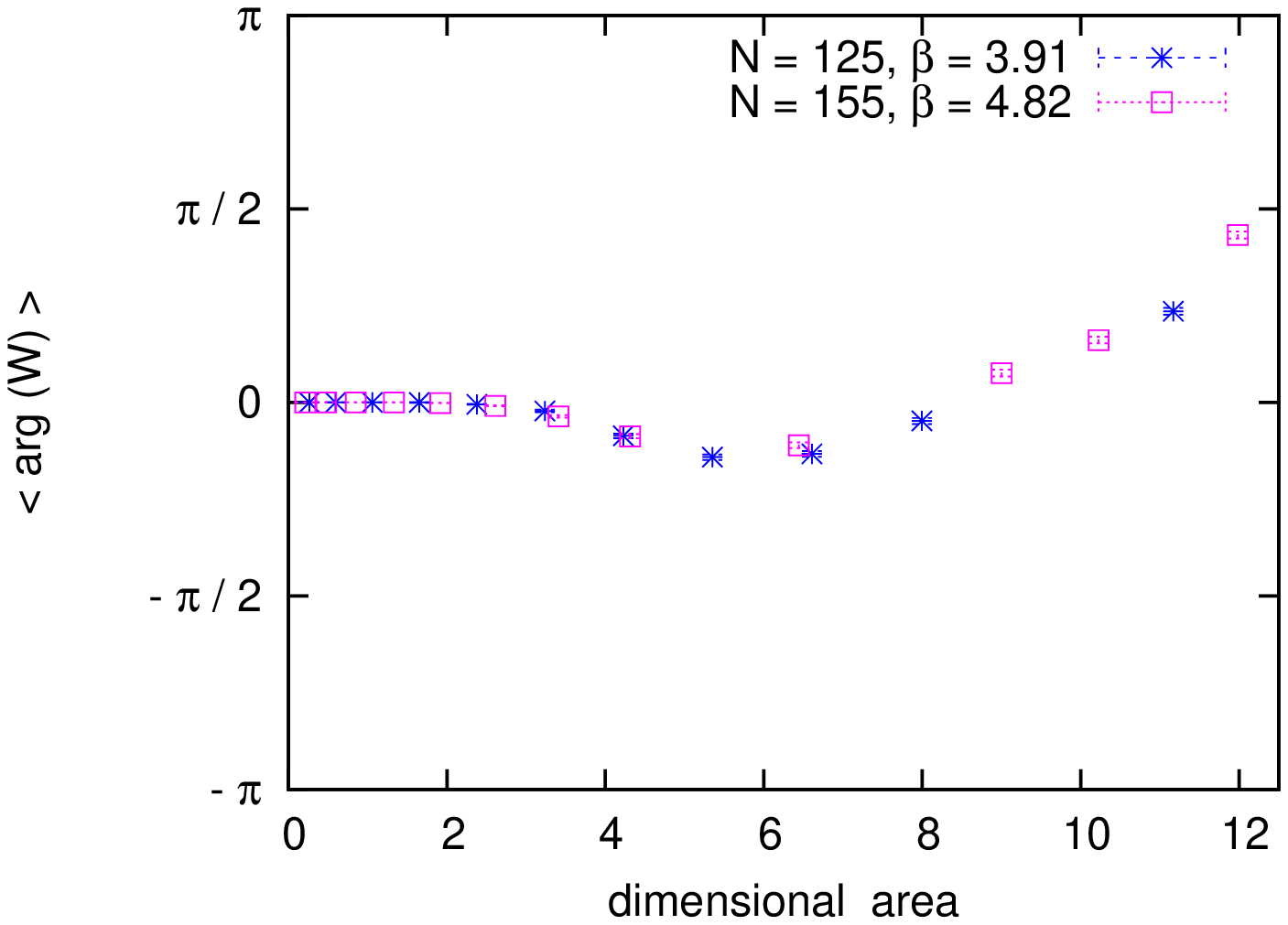}
\vspace*{-3mm}
  \caption{\emph{The phases of narrow rectangles with the shape 
$1 \times A$ at $\theta = 2.62$. We show the phase of 
$\langle W \rangle $ (on the left) and the expectation value
of $\, {\rm arg } (W) \, $ (on the right). Their values are very
similar, but these phases are not related to eq.\ (\ref{ABeq}). 
The essential observation
here is that the phases practically coincide for the two 
sets of parameters, which confirms again that we do see the
window that extrapolates to the DSL.}}
\label{rectphase}
\end{figure}

At last we consider the phases of the rectangles of the form
$1 \times A$, i.e.\ of an infinitesimally narrow shape in the
DSL. We mentioned in Section 4 that these phases do not follow
eq.\ (\ref{ABeq}). In Figure \ref{rectphase} we add the phases 
for these rectangles at $\theta = 2.62$. We show
$\, {\rm arg} ( \langle W \rangle ) \, $ on the left, and 
$\ \langle {\rm arg} (W) \rangle \, $ on the right. They
are very similar, and --- more importantly --- they are
in excellent agreement for the different $N$ values. This confirms
once more that we access the double scaling window 
with the parameters used.\\

This section presented additional APD symmetry breaking results,
now for $\langle |W| \rangle$ and $\langle {\rm arg}(W) \rangle$. 
The values and physical meanings of these observables are different
from those in Section 4, but qualitatively the results are in
full agreement.
The observables added here provide further strength
to our investigation of APD symmetry breaking --- see also the 
theoretical reasoning in the Appendix.

\section{The $SL(2,R)$ symmetry breaking}

While the breaking of the full APD symmetry has already been
demonstrated extensively, the specific case of the $SL(2,R)$ 
symmetry subgroup may seem less obvious from the Figures
shown so far. Hence this section focuses on rectangles only
to illustrate in particular the breaking of this subgroup.

\begin{figure}[ht!]
\begin{center}
\vspace*{2mm}
\includegraphics[width=.51\linewidth,angle=270]{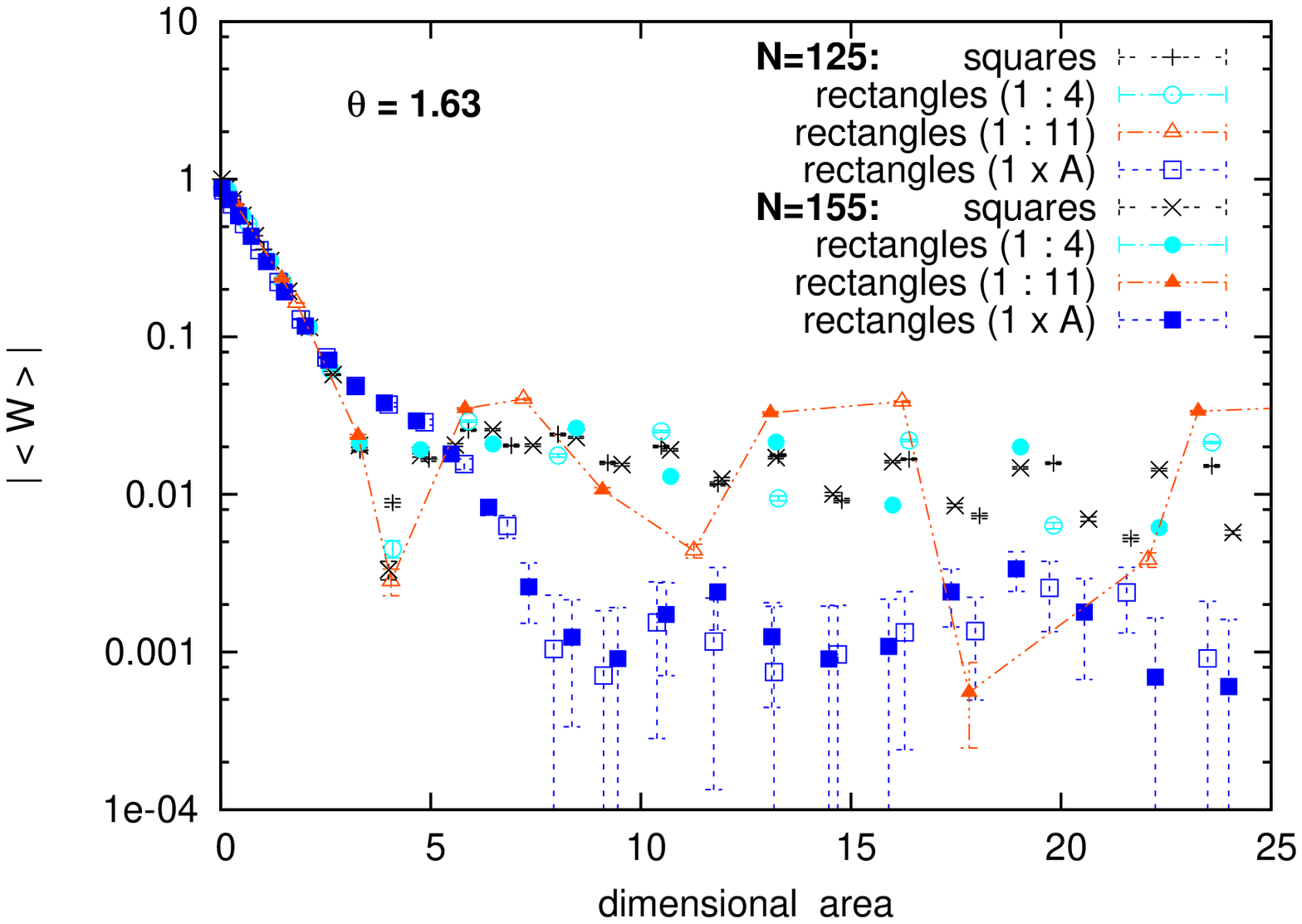}
\vspace*{2mm} 
\includegraphics[width=.51\linewidth,angle=270]{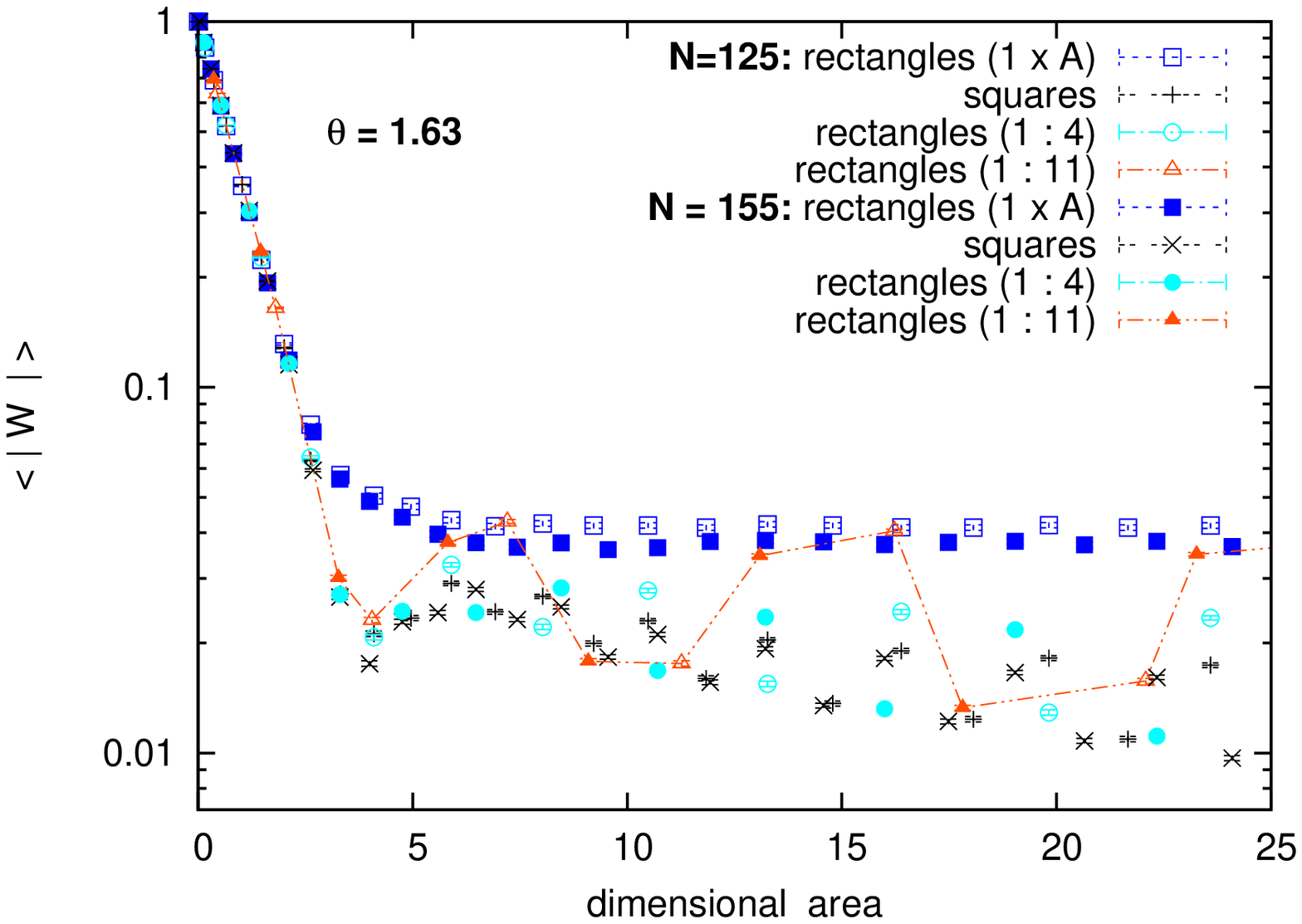}
\end{center}
\caption{\emph{The decay of the absolute values $|\langle W \rangle |$ 
(on top) and of $\langle |W| \rangle$ (below) 
for rectangular loops at $\theta = 1.63$. 
Beyond the area law regime, the rectangles with 
finite sides in the DSL (i.e.\ with a fixed side ratio) tend to 
oscillate around $\approx 0.01$ for $|\langle W \rangle |$ and a somewhat
larger value for $\langle |W| \rangle$. The amplitudes, however,
depend clearly on the side ratio, which shows the $SL(2,R)$
symmetry breaking. This is most evident for the rectangles with side 
ratio 11 (their data are connected by a line to guide the eye).}}
\label{decrec_theta1.63}
\end{figure}

\begin{figure}[ht!]
\begin{center}
\vspace*{2mm}
\includegraphics[width=.51\linewidth,angle=270]{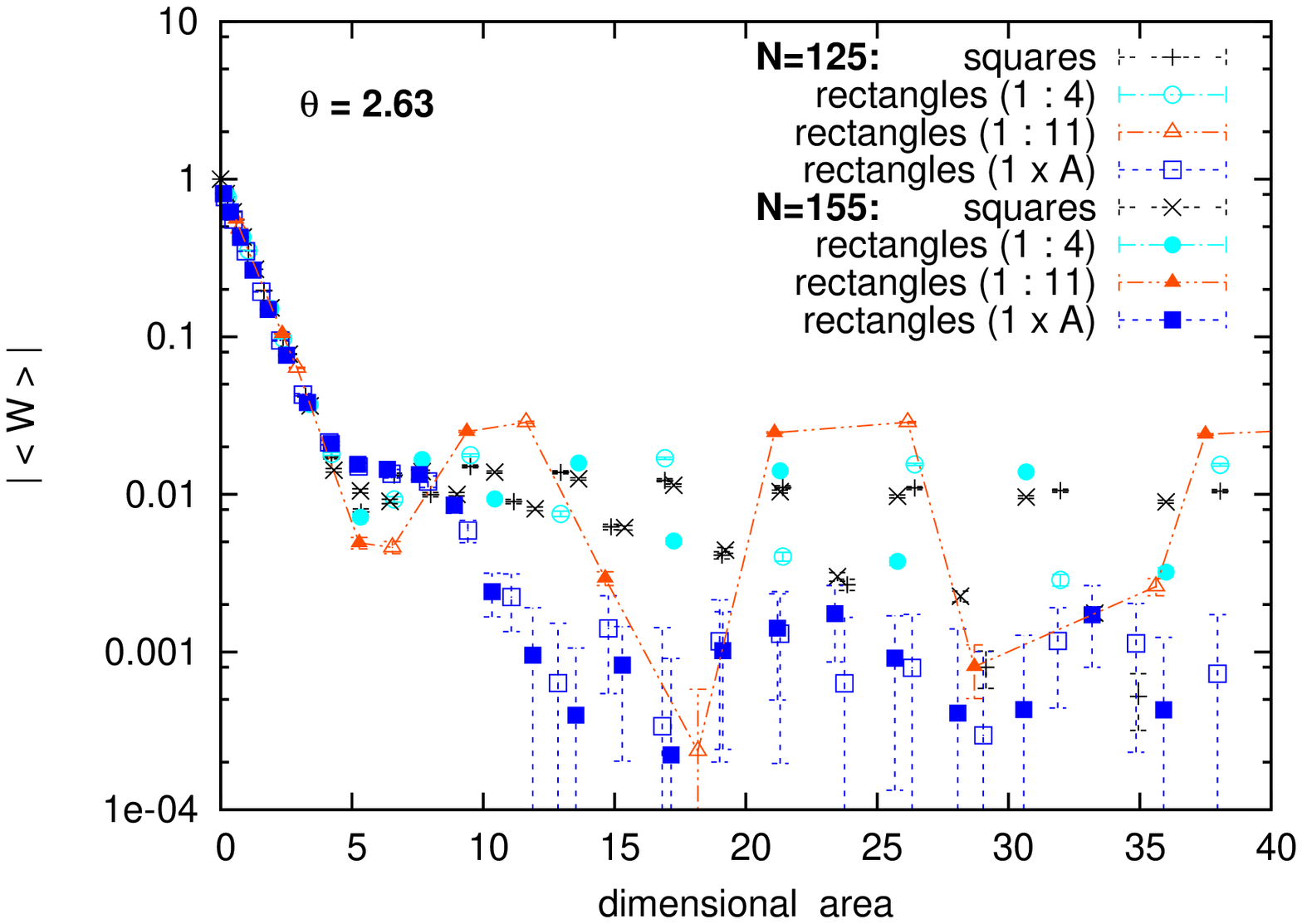}
\includegraphics[width=.51\linewidth,angle=270]{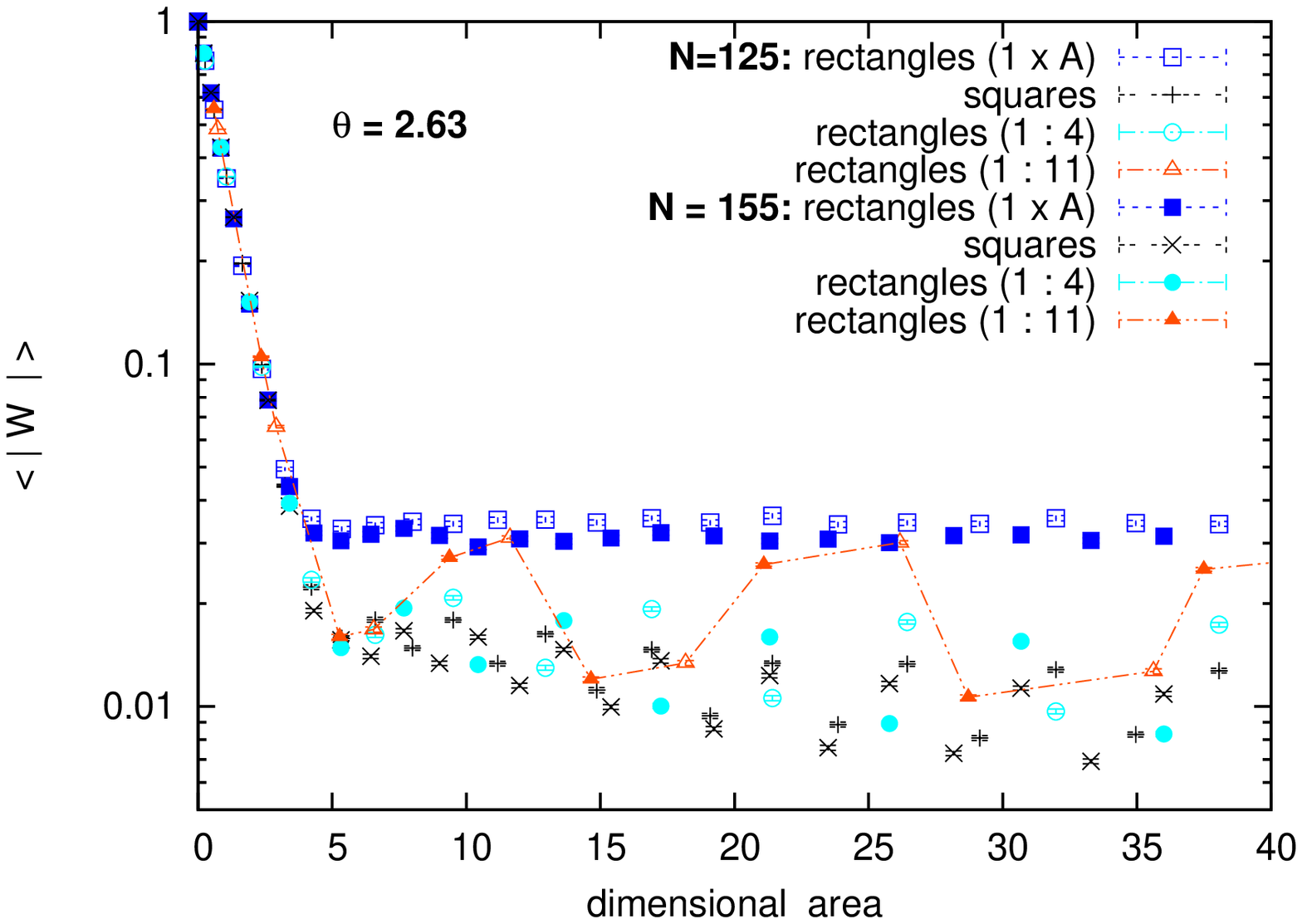}
\end{center}
\caption{\emph{The analogous plots to Figure \ref{decrec_theta1.63},
but now at $\theta = 2.63$. The observed decays of the rectangular
Wilson loops is very similar for both non-commutativity parameters
$\theta$ that we consider. Again the $SL(2,R)$ symmetry breaking can
been seen best from the data for rectangles with side 
ratio 11 (which are connected to guide the eye).}}
\label{decrec_theta2.63}
\end{figure}

Figures \ref{decrec_theta1.63} and \ref{decrec_theta2.63} 
are dedicated to the
decay of the absolute values $|\langle W \rangle |$ and
$\langle |W| \rangle$ for the rectangles. We show in both figures
the behaviour for fixed $\theta$, where we include the data from
$N=125$ and $N=155$ in the same plot. Here we also add results
for the decay of rectangles with a fixed side ratio of 11, which
are helpful to demonstrate the $SL(2,R)$ breaking more clearly.

For the rectangles with fixed 
shapes these observables tend to oscillate as we vary the dimensional
area beyond the area law regime. This observation is well compatible with 
the data from both $N$ values, hence it seems to characterise the DSL.
These oscillations have similar mean values, but the amplitude
is significantly larger for the ratio $11$ between the side length when
compared to ratio $4$ or $1$ (squares). 

We conclude that the $SL(2,R)$
symmetry is indeed more viable to  some approximation than the
rest of the APD symmetry group, but its breaking is nevertheless 
manifest on the non-perturbative level.

\section{Conclusions}

While APD symmetry holds in ordinary two-dimensional Yang-Mills 
theory, it turns out to be broken by non-commutativity.
Perturbation theory can reveal this effect, depending on the
order considered. Since this is still restrictive, to a finite 
order the symmetry may be protected in part \cite{noi}, with a
residual subgroup showing up in other approaches as well \cite{RicSza}.

However, on the non-perturbative level and by considering
finite $\theta$, the quantum effects unfold their full power
and destroy the APD symmetry, including the subgroup $SL(2,R)$. 
Here we presented explicit results for the APD symmetry breaking.
We considered four types of Wilson loops with polygonal
contours, without multiple-intersections.
At the same area, numerical simulations revealed the shape 
dependence of the Wilson loop expectation values on the lattice.
Our results were obtained for constant non-commutativity parameters
$\theta$ at different values of $N$ --- for increasing $N$ the volume
becomes larger (in dimensional units) and the lattice finer.
The results for the Wilson loops remain very stable, hence 
they allow for a reliable extrapolation to a continuous 
plane of infinite extent at fixed non-commutativity
(Double Scaling Limit). This limit reveals the breaking of
APD symmetry --- including the $SL(2,R)$ subgroup --- on the
non-perturbative level. Furthermore our results extend the loss
of APD symmetry to the discretised model in a finite volume.

To be more precise, the APD symmetry holds very well for
relatively small loop areas (in dimensional units). For a fixed 
lattice area this corresponds to the weak coupling regime, and 
hence the agreement with the leading order perturbative results
appears consistent. However, as the dimensional area increases,
this agreement --- and therefore the APD symmetry --- collapses:
it partially persists in the phase of the Wilson loops, but not
in their absolute values. We also observed that the $SL(2,R)$
subgroup has is a relatively high viability as an approximate
symmetry (which is consistent with perturbative results),
but its breaking becomes manifest for strongly anisotropic
transformations in this subgroup.

We conclude that NC gauge theory has a rich structure, 
even in $d=2$, far beyond its commutative counterpart. 
This conclusion rises hope for exciting effects.
On the other hand, it also means that an analytical
solution is unlikely. Therefore numerical results are of great
importance, as it is the case in commutative Yang-Mills theories
in four dimensions. \\

\noindent
{\small {\bf Acknowledgements: ~}
We are indebted to F.\ Hofheinz and J.\ Volkholz for
valuable help with the numerical work, 
and to A.\ Bassetto for reading the manuscript.
We also thank them, as well as
G.\ De Pol, H.\ Dorn, L.\ Griguolo, J.\ Nishimura, 
P.\ Sodano, Y.\ Susaki and F.\ Vian for interesting discussions.
The work of A.B.\ was supported by Istituto Nazionale di 
Fisica Nucleare (INFN), by the Deutsche Forschungsgemeinschaft 
(DFG), and by the Pan-European Research Infrastructure
on High Performance Computing (HPC-Europa). 
He thanks M.\ M\"{u}ller-Preu\ss ker for kindly supporting
his HPC-Europa application. The work of 
A.T.\ was supported by the DFG  within the Schwerpunktprogramm 
Stringtheorie 1096, by INFN with a Bruno Rossi postdoctoral fellowship, 
and by the U.S. Department of Energy (D.O.E.) under 
cooperative research agreement DE-FG02-05ER41360. 
The computations were
performed on the Hochleistungs-Rechenzentrum in Stuttgart
(HLRS) and on a PC cluster at Humboldt-Universit\"{a}t.
}

\appendix

\section{The perturbative treatment of $\langle |W| \rangle$}

In this appendix we comment on the perturbative 
treatment of the observable $\langle |W| \rangle$, along the lines 
of the formalism constructed in Refs. \cite{bnt1,bnt2}. 
We refer the reader to those works for the details and 
the relevant literature, while here we just recall the basic ingredients
needed for our argument. 

A convenient gauge choice for perturbative Wilson loop 
calculations is the 
light-cone gauge $A_{-}=0$. Faddeev-Popov ghosts are known to decouple also 
in the NC case, and the two-dimensional 
Lagrangian in this gauge looks indeed
free.
The non-trivial information on the dynamics is encoded in the singular
behaviour of the propagator 
\beq
D_{++} = {\rm i} \, [k_{-}^{- 2}] \ .
\eeq
Two prescriptions for the pole are important, 
the Cauchy principal value method by 't Hooft,
and the prescription due to Wu, Mandelstam and Leibbrandt \cite{WML},
which employs the propagator
\beq
D_{++} = {\rm i} \, [k_{-} + {\rm i} \epsilon k_{+}]^{- 2} \ .
\eeq 
In the commutative case,
the latter prescription is genuinely perturbative, while the former is
able to produce results for the Wilson loops, which
take into account non-perturbative contributions, and it
yields exponentiation (area law). But in the NC case, the former 
prescription behaves very wildly and has been soon abandoned, while the 
latter gives sensible results at all the orders considered. Also, 
the rotation to the Euclidean version of the theory turned out 
to be useful. 

When examining the perturbative series of the Wilson 
loop $\langle W \rangle$, one notices that the NC phase 
factor intermingles non-trivially, 
in the non-planar diagrams, with the propagators in the momentum 
and contour integrations. The results are
themselves expansions, typically in $1/ \theta \, $;
the Wilson loop is analytic around $\theta = \infty$. 

The perturbative series for the observable  $\langle |W| \rangle$ looks
a priori quite different from the one of $|\langle W \rangle|$. 
The relevant expansion in the $U(1)$ case reads
\begin{eqnarray}
\label{loopert}
\langle |W|\rangle &=& \left\langle \ \Big|
\sum_{n=0}^\infty ({\rm i} \, g)^n \int_{0}^1 ds_1 \ldots
\int_{s_{n-1}}^{1} ds_{n} 
\, \dot{\xi}_{-}(s_1)\ldots \dot{\xi}_{-}(s_{n}) \right.
\qquad \qquad \nonumber \\
&& \left. \int d^2x \,  A_{+}(x + \xi(s_1))
\star\ldots\star A_{+}(x + \xi(s_{n})) \Big| \ \right\rangle \ . 
\end{eqnarray}
Clearly the presence of the modulus inside the
quantum average complicates the task of rearranging the expansion in terms
of the Green functions of the theory, for one has first to write it 
schematically as $\sqrt{(\sum ...) (\sum...)^*}$ and then perform
a Taylor series in $g$. 
The fact that the sums begin with a constant, followed by terms with 
an increasing number of fields, allows
to formally treat the series as analytic around $g =0$. 
At the end, one 
computes the vacuum expectation value, based on the Green functions
\begin{equation}
\langle 0 |{\cal{T}}  A_{+}(x + \xi(s_1))
\star\ldots\star    A_{+}(x + \xi(s_{2n})) |0\rangle \ .
\end{equation} 
Each term
can in principle be treated similarly to the standard case: one is left with 
multiple integrals of propagators and Moyal phases, in momentum space
and on domains which are 
simplexes in the contour parameters. These contributions can be visualised 
by attaching the propagator lines along the contour itself: different simplexes
correspond then to different crossing patterns of the propagators. 

In the simplest case of the ${\cal{O}}(g^4)$ computation, 
the momentum integrals 
can be carried out exactly using complex plane techniques and identities
of Bessel functions, and the remaining contour integration is then  
performed numerically. For the new observable, care must be taken in 
computing some of the integrals, because one encounters terms like
\beq
\label{dangerous}
\left\langle \ \Big( \int_0^1 ds_i \, \dot{\xi}_{-}(s_i) \int d^2x \, 
\exp[ \, {\rm i} \, p (x + \xi(s_i))] \, 
A_{+} (p) \, \Big)^2 \ \right\rangle \ , 
\eeq
which appear already at the order ${\cal{O}}(g^2)$. This order 
is unaffected by the Moyal phase, but still the overall integration over the 
space-time base-point $x$ forces the appearance of a correlator 
$\langle A_{+} (0) A_{+} (0)\rangle$ at zero momentum which is infrared 
divergent. 
On the other hand, the integration over the closed contour vanishes because it 
reduces to $\int_0^1 ds_i \, \dot{\xi}_{-}(s_i) = 0$. Therefore,
one has to fix the order of integrations, such that 
the contour integration 
is performed first. Still, it is not clear if this can be made consistent 
at any order, or if other potentially divergent terms will appear due to 
the nesting of contour integrals, or, most important, due to interference 
with the NC phase factor.
We leave this interesting analysis for future developments. Here we would 
like to remark that, \it provided one can consistently treat these 
ambiguities, \rm then 
the arguments about order-by-order APD invariance at $\theta=0$
seem to apply. This implies that the observed shape-dependence 
of $\langle |W|\rangle$
in the NC plane appears as a valid argument for concluding that 
APD symmetry breaks down when 
going from $\theta=0$  to a finite $\theta$ value.

\end{document}